\documentclass[nopreprintline,final,12pt]{elsarticle}

\usepackage{amssymb}
\usepackage{amsmath}
\usepackage{xcolor}
\usepackage{tabularx}
\usepackage{booktabs}
\usepackage{algorithm2e}
\usepackage[caption=false,font=footnotesize]{subfig}

\newcommand{\boldv}{\mathbf{v}}
\newcommand{\boldw}{\mathbf{w}}
\DeclareMathOperator*{\argmin}{argmin}

\begin{document}

\begin{frontmatter}



\title{High-Fidelity RF Mapping: Assessing Environmental\\Modeling in 6G Network Digital Twins}


\author[label1]{Lorenzo~Cazzella}
\author[label1]{Francesco~Linsalata}
\author[label2]{Damiano~Badini}
\author[label1]{Matteo~Matteucci}
\author[label1]{Maurizio~Magarini}
\author[label1]{Umberto~Spagnolini}

\affiliation[label1]{organization={Politecnico di Milano, Milan, Italy}}
\affiliation[label2]{organization={Huawei Technologies Italia S.r.l., Segrate, Italy}}

\begin{abstract}
The design of accurate Digital Twins (DTs) of electromagnetic environments strictly depends on the fidelity of the underlying environmental modeling. Evaluating the differences among diverse levels of modeling accuracy is key to determine the relevance of the model features towards both efficient and accurate DT simulations. In this paper, we propose two metrics, the Hausdorff ray tracing (HRT) and chamfer ray tracing (CRT) distances, to consistently compare the temporal, angular and power features between two ray tracing simulations performed on 3D scenarios featured by environmental changes. To evaluate the introduced metrics, we considered a high-fidelity digital twin model of an area of Milan, Italy and we enriched it with two different types of environmental changes: (i) the inclusion of parked vehicles meshes, and (ii) the segmentation of the buildings facade faces to separate the windows mesh components from the rest of the building. We performed grid-based and vehicular ray tracing simulations at 28 GHz carrier frequency on the obtained scenarios integrating the NVIDIA Sionna RT ray tracing simulator with the SUMO vehicular traffic simulator. Both the HRT and CRT metrics highlighted the areas of the scenarios where the simulated radio propagation features differ owing to the introduced mesh integrations, while the vehicular ray tracing simulations allowed to uncover the distance patterns arising along realistic vehicular trajectories.
\end{abstract}




\end{frontmatter}



\section{Introduction} \label{sec1}

The accurate modeling of electromagnetic (EM) environment has significant implications for improving key parameters in network design and deployment \cite{7509384}. Within this context, the concept of the Digital Twin (DT) emerges, representing a sophisticated but interesting approach to constructing precise digital replicas of physical entities or processes. These DTs provide real-time insight into their corresponding physical counterparts, effectively integrating simulations with existing databases \cite{DTMagazine}. This framework necessitates the development of appropriate models for the continuous update and enhancement of the virtual representations, ensuring their matching with real-world entities. Achieving this correspondence enables a closed-loop system in which decisions concerning the physical entity are informed by real-time data derived from the DT \cite{Pegu2505:Toward}.

The recent emergence of network virtualization through the Open Radio Access Network (ORAN) architecture \cite{polese2023understanding} complements these advancements by providing the necessary network data for constructing accurate DTs. This approach alleviates data management and exchange burdens on network nodes, while enabling ORAN-based applications to leverage DTs for comprehensive 6G scenario inference without incurring additional communication overhead with the infrastructure \cite{moro2023advancing}.

\subsection{Related work}

Digital Twins (DTs) have been extensively explored across various domains, including real-time monitoring and control of industrial systems, risk assessment in transportation, and smart scheduling for urban infrastructure. In wireless communications, research has primarily addressed operational challenges such as edge computing, network optimization, and service management \cite{DTMagazine}. These studies underscore the transformative potential of DTs to improve the efficiency, adaptability, and performance of communication systems by leveraging real-time digital representations.

A key factor in these advancements is the integration of multi-modal data, which accelerates decision-making processes in communication systems. Traditional tasks like beam prediction and localization often involve exhaustive searches over extensive candidate sets, leading to significant delays \cite{gu2022multimodality}. To overcome these challenges, Gu \textit{et al.} introduced a dataset specifically designed for beam selection in vehicular networks operating in millimeter-wave (mmWave) bands \cite{gu2022multimodality}. While these datasets streamline decision-making for specific scenarios, our work advances this approach by focusing on real-time DTs that model physical environments and simulate wireless signal propagation.

Despite their accuracy, ray-based simulations have historically faced substantial computational demands, limiting their feasibility for real-time applications \cite{zhu2024toward}. However, recent advancements---including adaptive ray launching, simplified urban modeling, and computationally efficient frameworks---have addressed these challenges \cite{9459462}. Ray-based propagation simulations remain a powerful tool for wireless channel modeling, offering precise parameter estimates, such as path loss, angles of arrival and departure, propagation delays, and Doppler shifts. These simulations utilize high-frequency approximations of Maxwell’s equations and model electromagnetic wave propagation using rays \cite{7152831}. When combined with high-resolution 3D urban maps, they deliver accurate representations of complex propagation environments, particularly in dense urban areas.

The integration of ray-based simulation engines into DT-enabled systems has further refined real-world electromagnetic environment modeling. By simulating phenomena like reflection, diffraction, and scattering, these engines generate high-fidelity representations of urban landscapes, which are critical for modeling high-frequency signals in emerging sixth-generation (6G) networks. These detailed radio maps enable advanced applications involving connected vehicles, drones, and other networked entities \cite{10198573, zhu2024toward, sionna}.

Real-time performance in ray-based simulations remains a challenge, particularly in dynamic environments such as vehicular networks, where conditions evolve more rapidly than simulations can adapt \cite{CazzellaVTC}. To address this, researchers have developed various optimizations, including simplified ray-based modeling for end-to-end network simulations \cite{9459462}, dynamic computational methods \cite{10089404}, adaptive ray launching algorithms for urban scenarios \cite{8529268}, and streamlined building representations \cite{10238157}.

One example is the Boston Digital Twin, which integrates high-fidelity 3D city models with optimized computational techniques to enable ray-based simulations within DT systems \cite{BostonTWIN}. Despite these advancements, further improvements are needed for highly dynamic and complex channel modeling scenarios, such as those encountered in 6G networks. Progress in integrating ray-based simulations into DT-enabled systems remains vital for achieving precise, real-time digital representations applicable across diverse domains of wireless communications.

While scene geometry acquisition has become relatively straightforward, the characterization of material properties continues to pose significant challenges, requiring accurate calibration through channel measurements. Ray tracer calibration methods can be grouped into three main approaches, each with limitations. The first involves direct measurement of material properties for individual objects \cite{10705152}, which does not scale well to complex scenes or resolve ray tracer-specific mismatches. The second adjusts material parameters to match aggregate statistics, such as path loss or delay spread \cite{kanhere2023calibration, he2017channel}, but oversimplifies the environment and sacrifices granularity. The third method aligns individual rays from measurements to ray tracing predictions for per-ray calibration \cite{charbonnier2020calibration}, but is computationally intensive, particularly for large datasets.

Recent research has also explored machine learning (ML) surrogates for ray tracers, trained on measured data to predict metrics such as path loss or line of sight condition \cite{zhu2025exploiting}. These methods complement advancements in DT systems by enhancing the accuracy of ray-based simulations. Notably, \cite{bakirtzis2022deepray} introduced a gradient-based calibration approach with differentiable parametrization for material properties, scattering, and antenna patterns, addressing several limitations of traditional techniques.

Material segmentation precision is equally critical, as neglecting it can result in highly reflective paths and system instability, underscoring the importance of accurate modeling. For instance, \cite{10556262} introduced WiSegRT, an open-source dataset for indoor radio propagation modeling, generated using a differentiable ray tracer applied to segmented 3D indoor environments. Similarly, mmSV \cite{kamari2023mmsv} reconstructs 3D environments from street view images, assigning materials through semantic segmentation. Other works, such as \cite{levie2021radiounet}, utilize 2D layouts to train neural networks for radiomap predictions, while \cite{kocevska2023identification} extends this to basic 3D environments, predicting radiomaps and detailed channel metrics such as CIR, AoA, and AoD.

\subsection{Contributions and outline}
In this paper, we propose the following contributions:
\begin{itemize}
    \item We introduce the Hausdorff ray tracing (HRT) and chamfer ray tracing (CRT) distance metrics to compare two sets of delay, power, DoA and DoD ray tracing parameters interpreting them as two point clouds embedded in a linear space of suitable dimension. We use the proposed metrics to perform the comparison between a plain DT scenario---composed only by buildings and ground meshes---and the scenarios enriched with the discussed vehicular and windows mesh features.
    
    \item We extend a high frequency EM digital twin of an urban environment produced considering the detailed modeling of two environmental features: (i) the presence of meshes for the parked vehicles distributed in the environment from usual positions derived from the observation of the physical twin; (ii) the segmentation of the buildings meshes to separate the windows from the remaining mesh and assign to them a different radio material.

    \item We assess the impact of the presence of parked vehicles and accurately segmented buildings windows through a comparative analysis of channel simulations by using the introduced HRT and CRT metrics. The experiments are performed by independently comparing a base urban scenario with (i) the integration of parked vehicles' meshes, and with (ii) the segmentation of windows on the buildings' facades. We achieved accurate ray tracing simulations over the considered scenarios using the NVIDIA Sionna RT ray tracing simulator. We analyze the distance in the spatial, temporal and power domains, focusing on both grid-based fingerprinting simulations and on vehicular simulations obtained integrating NVIDIA Sionna RT with the SUMO vehicular traffic simulator.
\end{itemize}

The paper is organized as follows: Section \ref{sec:dt_modeling} details the construction of the considered urban DT model, specifying the considered scenario, the modeling and addition of parked vehicles' meshes, and the segmentation of the buildings' windows for consistent radio materials matching. Section \ref{sec:method} introduces the proposed method to compare ray tracing simulations under environmental changes by means of the Hausdorff (HRT) and Chamfer (CRT) ray tracing distances over a set of parameters directly derived from the propagation paths obtained through a ray tracing simulation. In Section \ref{sec:results}, we provide the simulation results, investigating the use of the proposed metrics to assess the differences over environmental changes in parked vehicles and buildings' windows modeling. Finally, Section \ref{sec:conclusion} draws the conclusions.

\section{Digital Twin Modeling}\label{sec:dt_modeling}

In this section, we detail the DT model adopted in this work, along with the procedures that we employed to enrich the environment with the inclusion of models of parked vehicles, discussed in Section \ref{sec:parked_vehicles_modeling}, and with the segmentation of the buildings meshes to separate the windows components in order to assign two different radio-materials to achieve higher fidelity, as outlined in Section \ref{sec:windows_segmentation}.

\subsection{Considered scenario}\label{sec:scenario}

\begin{figure*}[t!]
    \centering
    \subfloat[][OSM scenario top view]{\includegraphics[width=.443\textwidth]{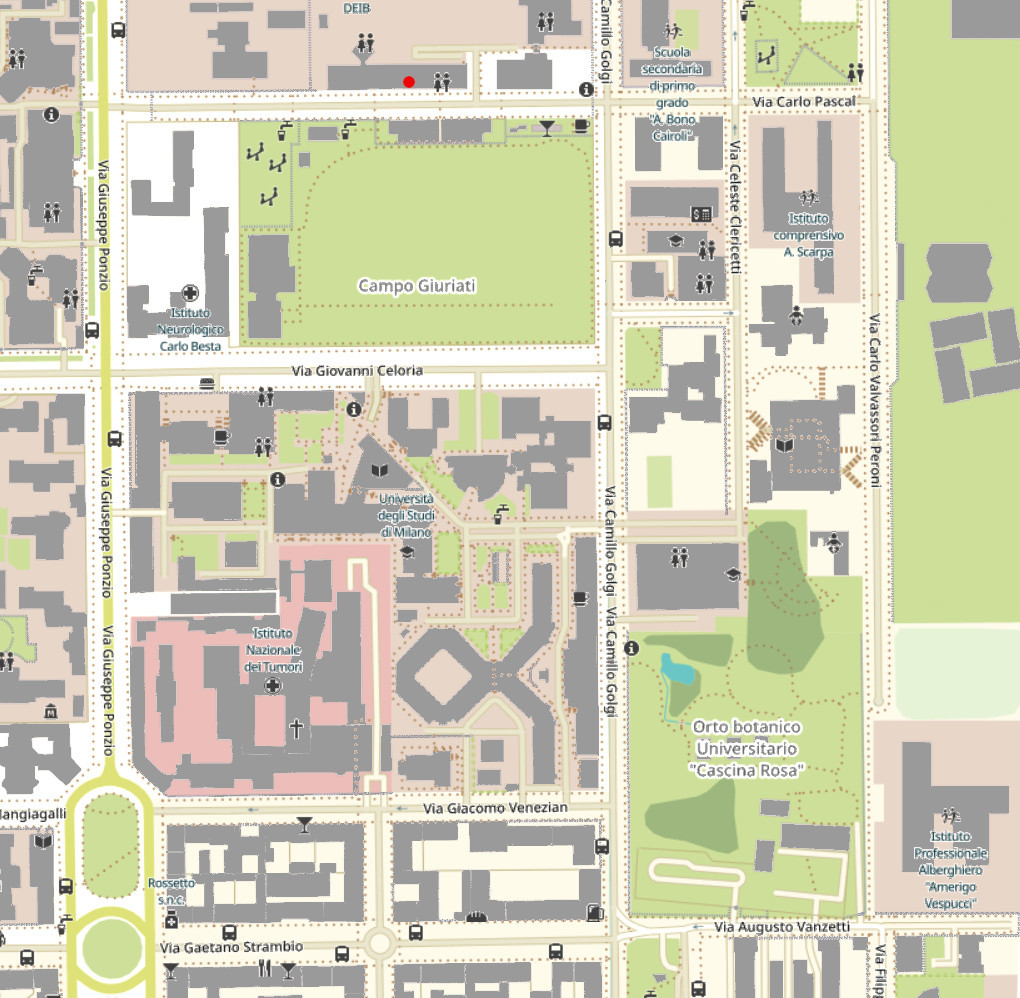}}
    \hspace{0.1cm}
    \subfloat[][Scenario 3D meshes front view]{\includegraphics[width=.52\textwidth]{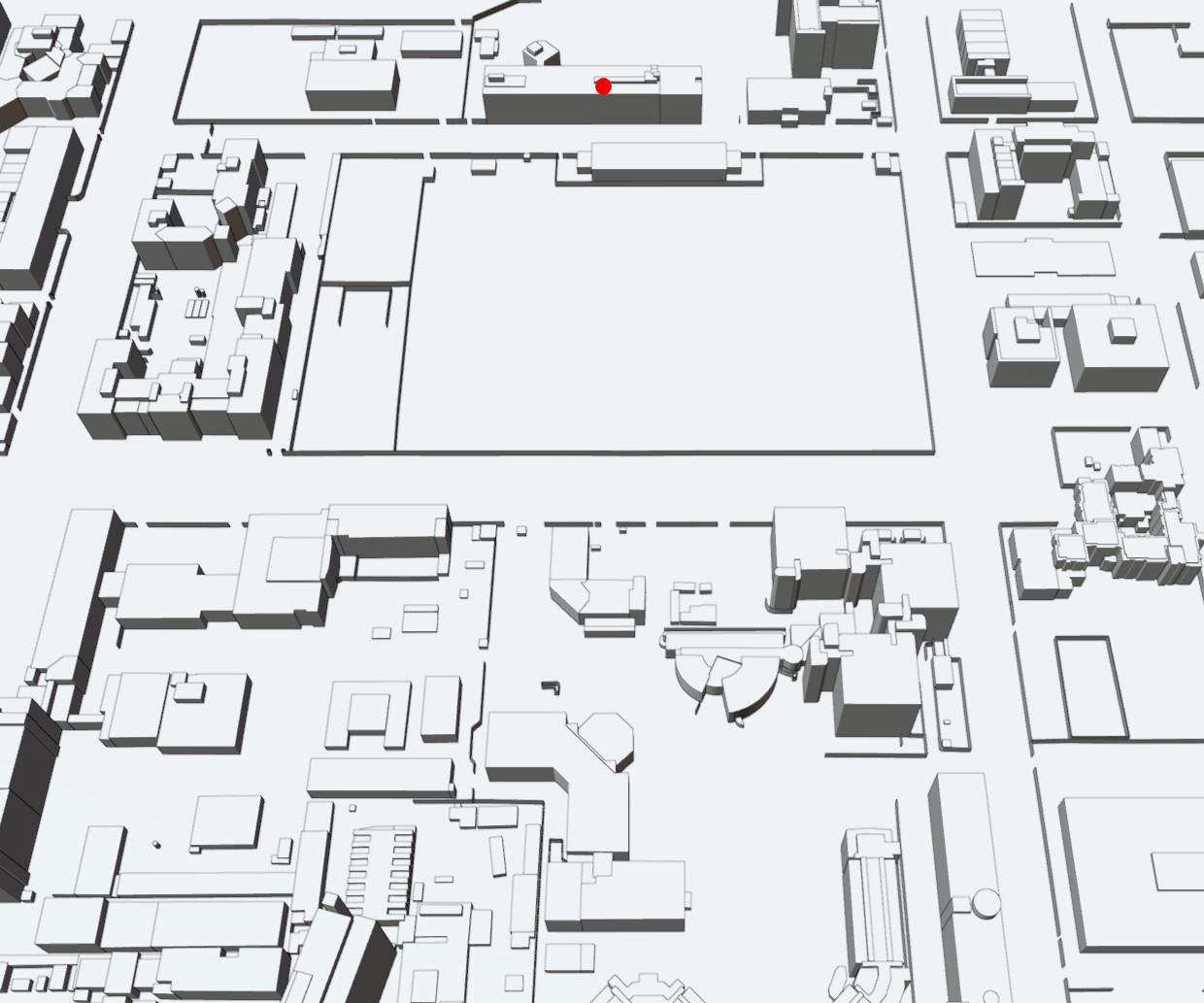}}\\
    \caption{Considered DT scenario of an area of Milan, Italy. (a) presents a top view of the scenario gathered from OpenStreetMap, while (b) shows a frontal view of the scenario meshes. The red point indicates the position of the BS considered for I2V ray tracing simulations.}
    \label{fig:scenario}
\end{figure*}

In this work, we consider a high-fidelity 3D model of an area of Milan, Italy surrounding the Department of Electronics Information and Bioengineering (DEIB) of Politecnico di Milano. The model includes buildings and walls and has an extension of $550 \times 670$ m. Its 3D meshes are depicted in Fig.~\ref{fig:scenario}, which shows on the left a top view of the urban setting from OpenStreetMap, and on the right a front view of the scenario meshes. The buildings and wall meshes are modeled as a set of stacked irregular right prisms, which suitably represents and effectively approximates the mostly planar buildings facades and walls structures.

To enable ray tracing simulation, the scenario has been preprocessed to include a ground and to specify the radio materials for the meshes. Finally, the meshes have been exported for the integration with the ray tracing simulator. To allow the integration of the selected scenario with vehicular traffic simulations, we gathered from OpenStreetMap \cite{openstreetmap} the traffic network topology information for the same area. The availability of the road topologies data enabled the simulation of realistic VEs positions to enable the comparison of the different ray tracing simulations across common vehicular urban conditions. The steps followed for the integration of the scenario within the vehicular traffic and ray tracing simulators is detailed in Section \ref{sec:simulation_setup}.

\subsection{Modeling parked vehicles meshes}\label{sec:parked_vehicles_modeling}
\begin{figure*}[t!]
    \centering
    \subfloat[][Sedan]{\includegraphics[width=.32\textwidth]{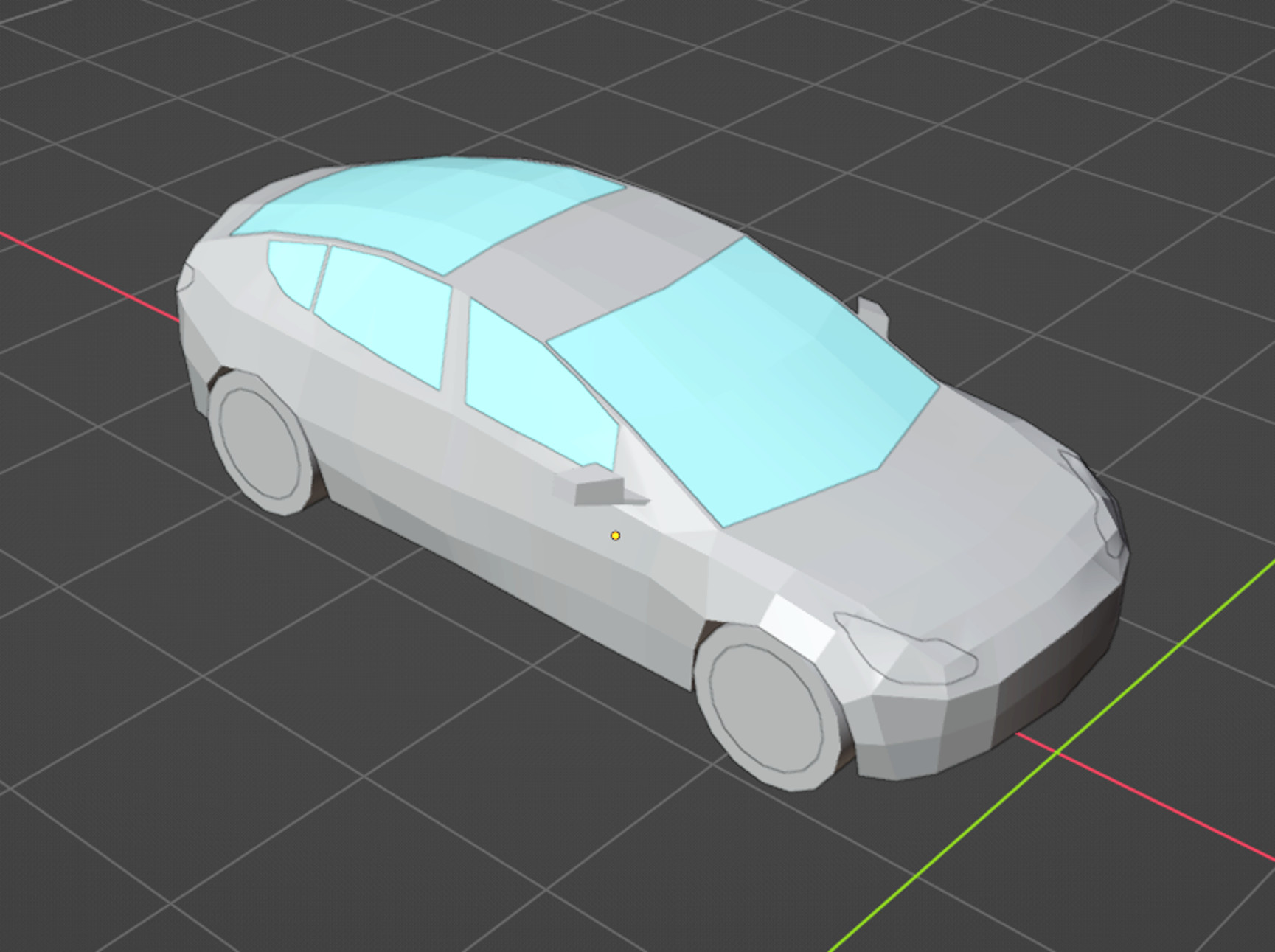}}
    \hspace{0.1cm}
    \subfloat[][Hatchback]{\includegraphics[width=.32\textwidth]{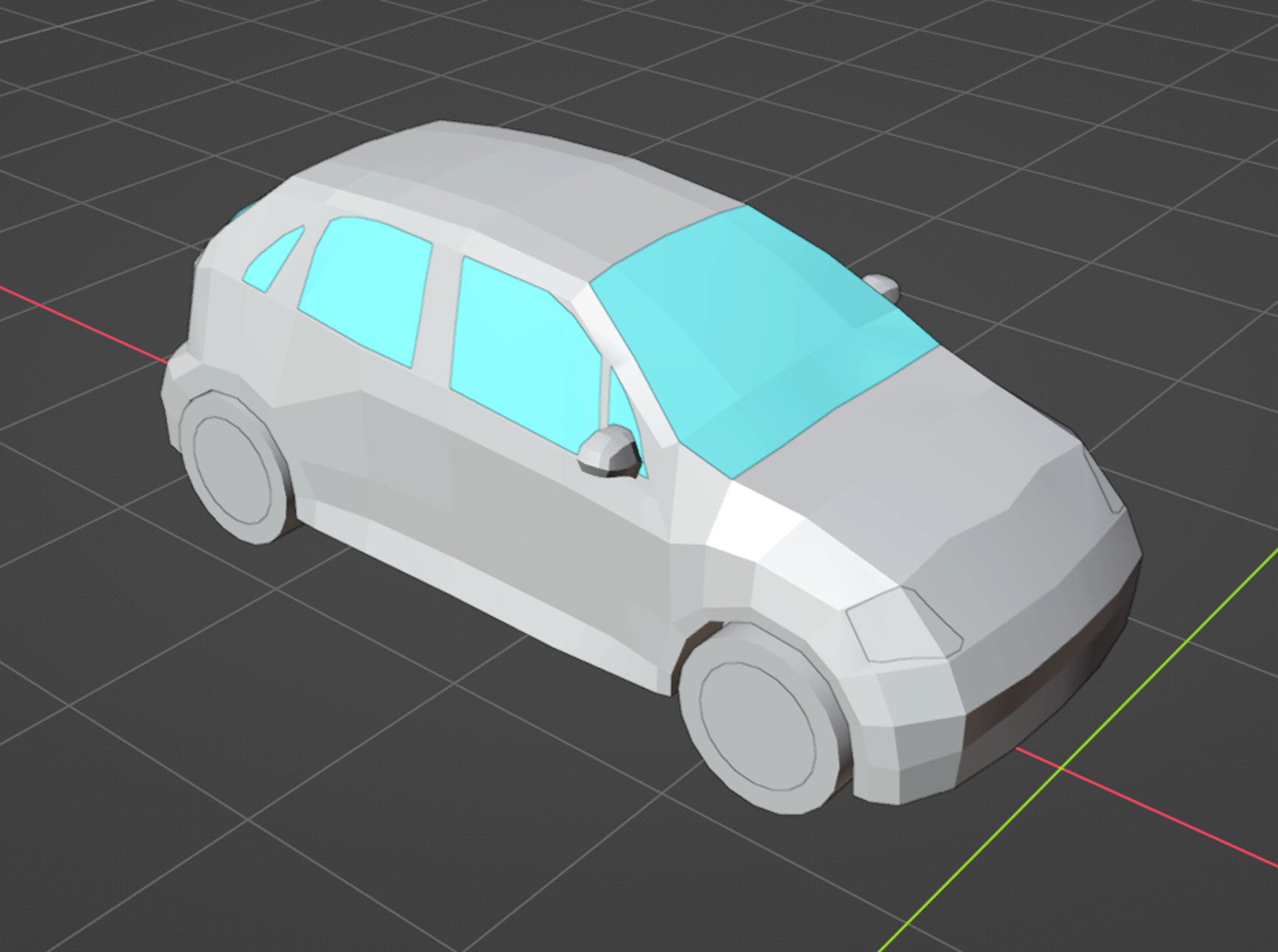}}
    \hspace{0.1cm}
    \subfloat[][Truck]{\includegraphics[width=.315\textwidth]{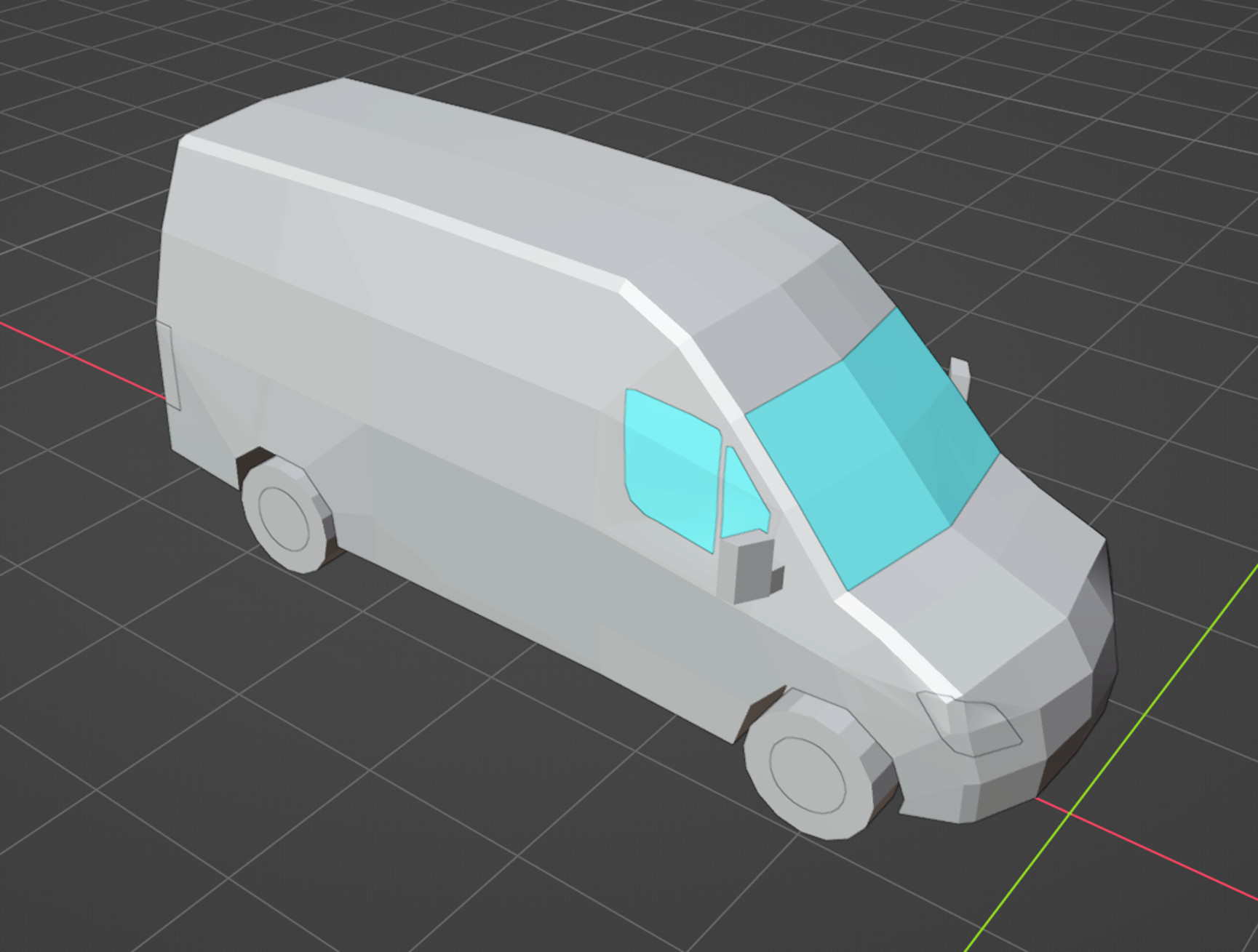}}\\
    \caption{Considered vehicle meshes for parked vehicles modeling. The depicted models have been selected among the ones available in the CARLA automotive simulator and have been manually segmented to separate the windows mesh components.}
    \label{fig:parked_vehicles_meshes}
\end{figure*}

In this section, we describe the procedure that we adopted to model the presence of parked vehicles in the reference urban scenario. We considered three different types of vehicle meshes to model the presence of parked vehicles. We selected the meshes among the vehicular assets available in the CARLA automotive simulator \cite{Carla} for their level of detail and openness. Among them, we have chosen the Tesla Model 3, Citroën C3, and Mercedes Sprinter models, which in the following we will indicate as sedan, hatchback and truck, respectively. The corresponding meshes of the three selected vehicle types are depicted in Fig. \ref{fig:parked_vehicles_meshes}. This set of vehicle models has been selected owing to its variety and as consistent approximation of the meshes of vehicles commonly used in urban environments. Owing to the high number of vehicle models to be set in the reference scenario, and to the spatial extension of the latter, we chose the version of the CARLA vehicle models depicted in Fig. \ref{fig:parked_vehicles_meshes} instead of the most detailed ones as a trade-off between level of detail and mesh complexity. This choice has been driven by the use of the final integrated model within NVIDIA Sionna \cite{hoydis2022sionna} for ray tracing simulation.

\begin{figure*}[t!]
    \centering
    \subfloat[][Scenario with parked vehicles.]{\includegraphics[width=.52\textwidth]{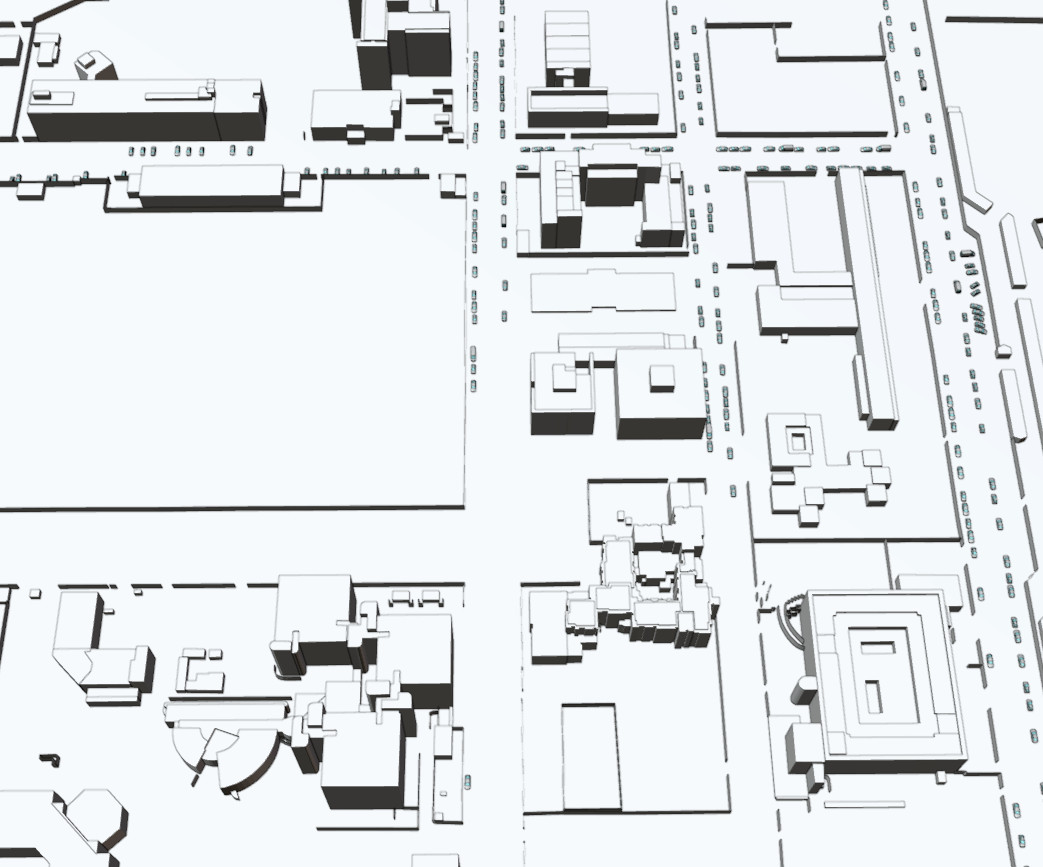}}
    \hspace{0.1cm}
    \subfloat[][Parked vehicles detail.]{\includegraphics[width=.46\textwidth]{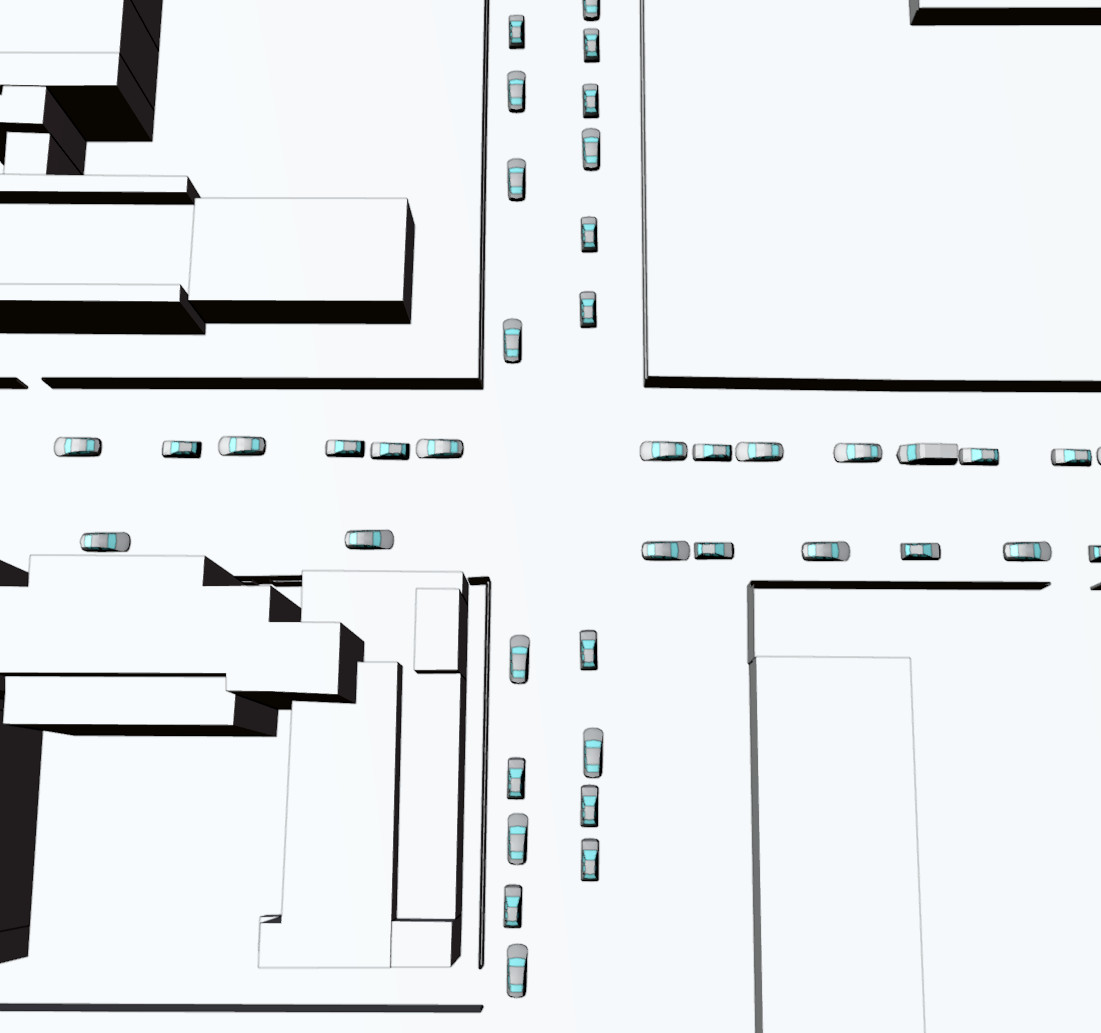}}
    \hspace{0.1cm}
    \caption{Base scenario enriched with parked vehicles. (a) shows a front view of the scenario, while (b) depicts a detail of parked vehicles at one of the intersections.}
    \label{fig:scenario_parked_detail}
\end{figure*}

\textbf{Vehicle meshes segmentation}. To enhance the fidelity of the model to real-world scenarios, the 3D meshes of the chosen vehicle types have been manually segmented to differentiate the areas related to the windows with respect to the rest of the vehicle body and the remaining components. The segmentation has been performed in Blender, keeping the overall mesh closed and preserving the orientation of the segmented faces. Fig. \ref{fig:parked_vehicles_meshes} shows an example of such a segmentation, where the segmented windows are highlighted in a different color with respect to the vehicle body.

\textbf{Radio material assignment}. To enable the vehicle meshes for ray tracing simulation in NVIDIA Sionna RT, the windows and body segmented components have been assigned to the glass and perfect electric conductor (PEC) radio materials, respectively, using the material parameters recommended by ITU in \cite{ITU-R-2040} corresponding to the adopted 28 GHz carrier frequency.

\textbf{Parked vehicles positioning}. To achieve realistic positioning of the vehicles, we set the vehicles locations and orientations according to satellite imagery retrieved from Google Maps \cite{gmaps} for the streets in the reference area. The satellite imagery has been mapped into the Blender 3D model of the scenario and the vehicle meshes have been selected depending on the corresponding vehicle type and superposed in top view. We totally included 505 vehicle meshes, divided into 331 sedans, 154 hatchbacks and 20 trucks. Fig. \ref{fig:scenario_parked_detail} depicts two views of the base scenario enriched with parked vehicle models in common positions as observed from satellite imagery of the area.

\subsection{Buildings windows segmentation for consistent radio material matching}\label{sec:windows_segmentation}

\begin{figure}[t!]
    \centering
    \includegraphics[width=\linewidth]{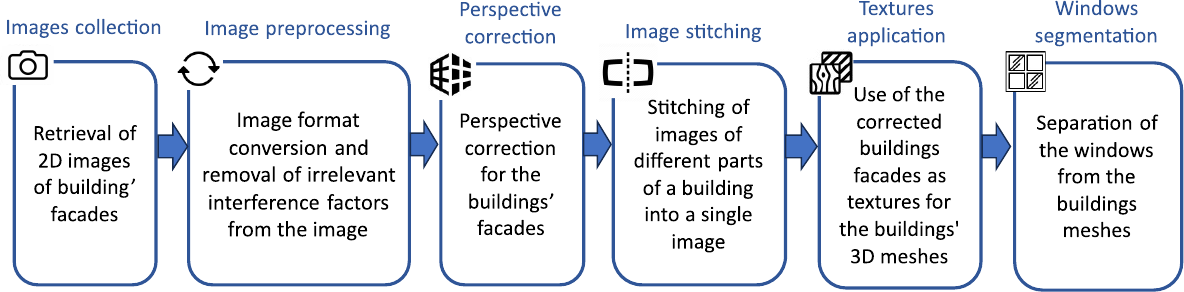}
    \caption{Windows segmentation pipeline.}
    \label{fig:segmentation_pipeline}
\end{figure}

We provide here the details on the workflow followed to segment the windows in the buildings facades within considered the reference scenario. In the following, we discuss in detail the followed pipeline, which we provide in Fig. \ref{fig:segmentation_pipeline}. 

\textbf{Facade images retrieval}: Since only the buildings meshes without any texture were available, the first step towards meshes segmentation has been the retrieval of camera images of the buildings facades. A total of 616 images with a resolution of $4032\times 3024$ of the buildings in the scenario have been collected in an acquisition campaign across the considered area of Milan, covering all the buildings facades of interest.

\textbf{Image preprocessing}: The retrieved images have been preprocessed to adapt their format for the next stages. In the presence of irrelevant interference factors that would prevent from the identification of the windows boundaries---e.g., the presence of trees or vehicles in front of windows meshes---multiple images of the facade have been acquired in order to identify more easily the windows meshes from multiple independent views. The information from the multiple views has then been used in the image stitching process described below to recover an image for all the windows.

\textbf{Perspective correction}: The images of the buildings facades have been corrected through a homographic transformation to adapt them for the application to the planar buildings facades faces within the 3D model.

\textbf{Image stitching}: Some of the photographed buildings were too large to include the whole facade in a single image. In those cases, applied image stitching to produce a single corrected planar image of the facade starting from multiple acquired images.

\textbf{Textures application}: We applied the corrected and stitched images to the corresponding facade faces within the 3D meshes model using the Blender 3D modeling software. The image textures have been disposed to maximally match the available 3D model faces while remaining consistent with the real world scenario.

\begin{figure*}[t!]
    \centering
    \subfloat[][Segmented scenario.]{\includegraphics[width=.513\textwidth]{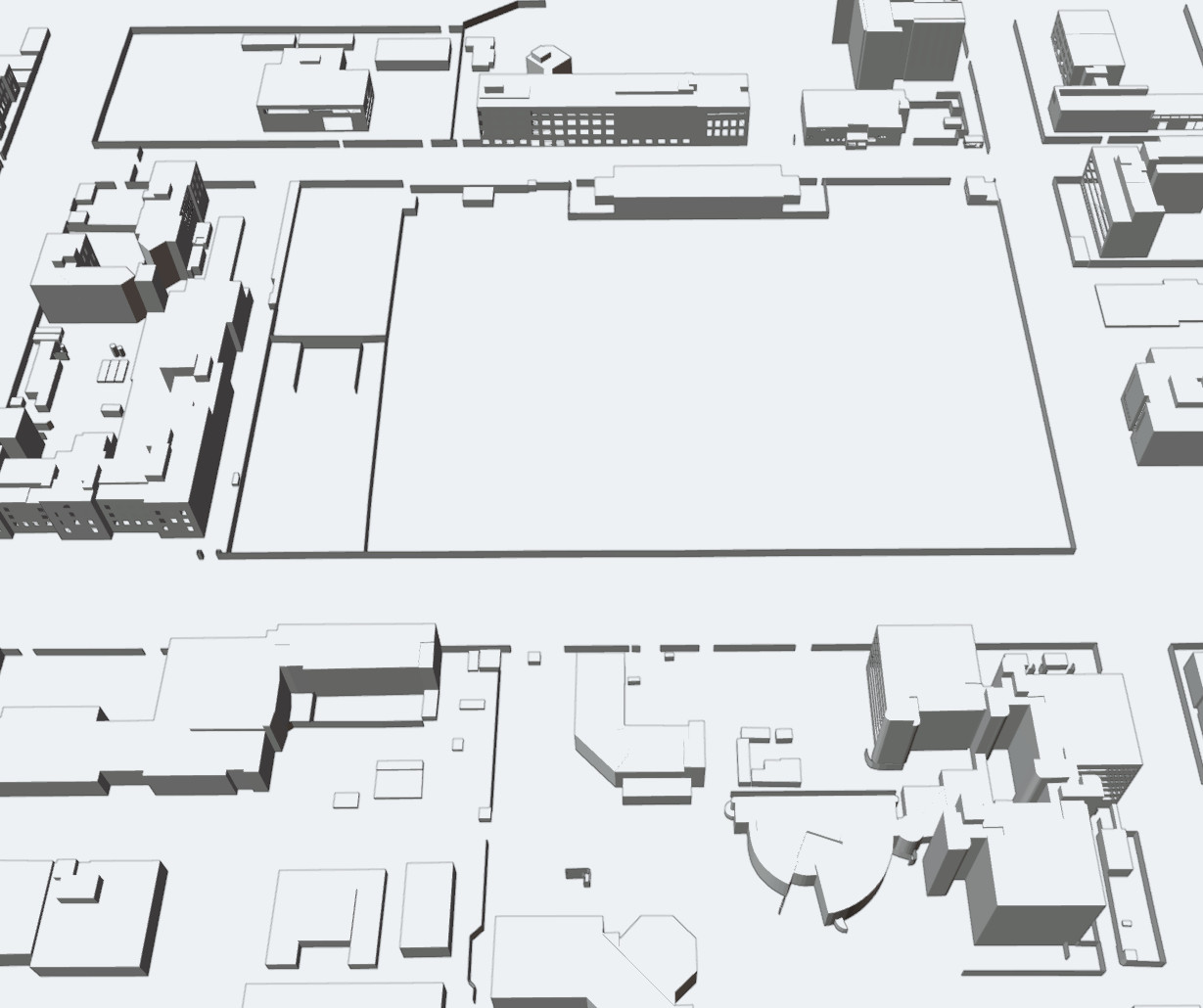}}
    \hspace{0.1cm}
    \subfloat[][Segmented building detail.]{\includegraphics[width=.46\textwidth]{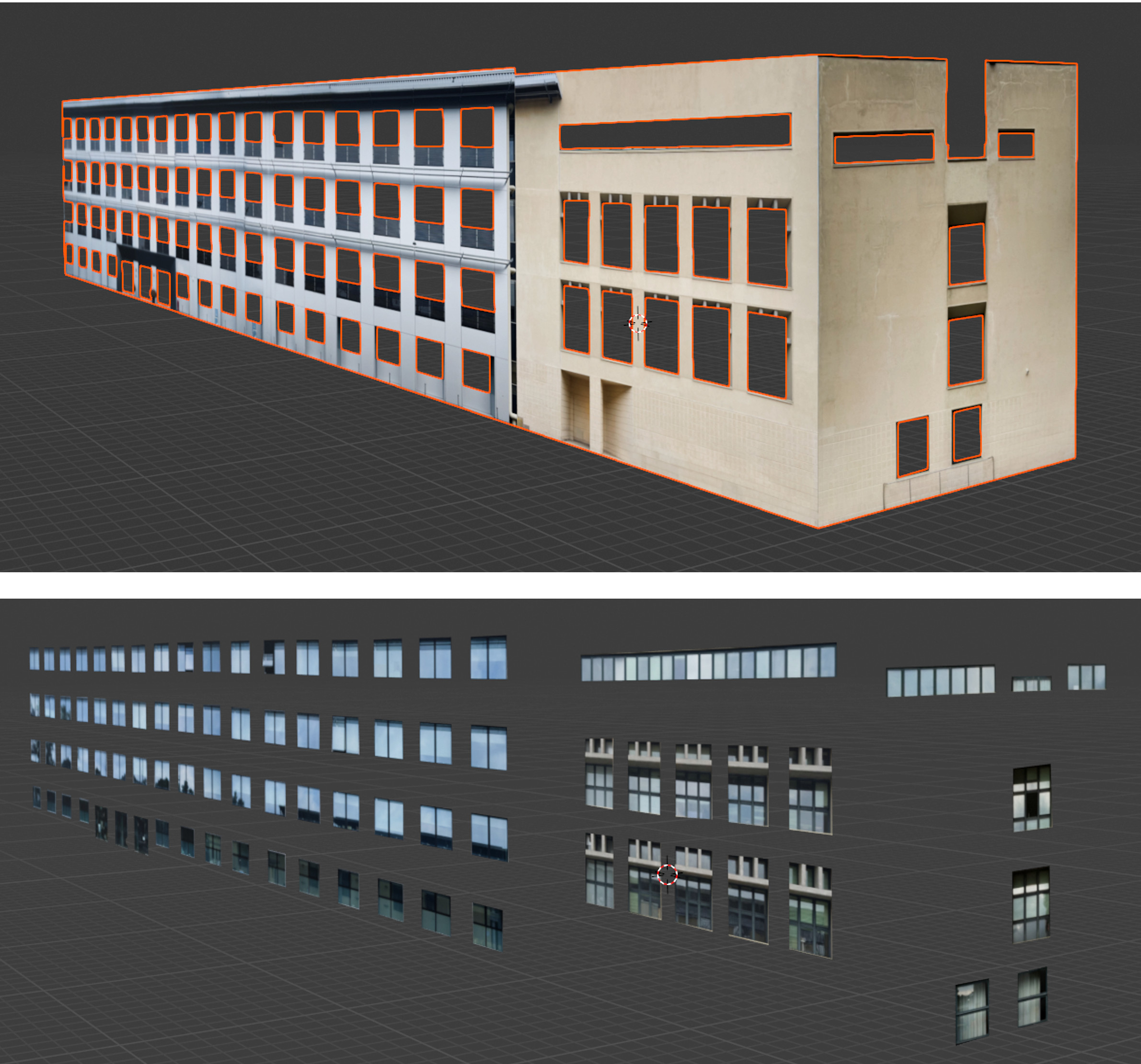}}
    \hspace{0.1cm}
    \caption{Base scenario enriched with building windows segmentation. (a) shows the segmented scenario, where the windows are missing from the buildings to highlight their segmentation, while (b) represents the detail of a segmented building, with the top image showing the building meshes without windows and the bottom image representing the opposite.}
    \label{fig:segmented_windows_scenario}
\end{figure*}

\textbf{Windows segmentation}: Finally, we manually segmented the windows mesh components from the buildings meshes using the Blender modeling tools using the applied textures as reference for the windows boundaries. The resulting meshes are depicted in Fig. \ref{fig:segmented_windows_scenario}, which shows a frontal view of the segmented scenario---where windows are missing to highlight the result of the segmentation procedure---and a detail of a building segmented using the textures applied to it from the corresponding acquired images.

To prevent the generation of different rays from the ray tracing simulator---which may result from the use of different sets of meshes produced by the windows segmentation---we employ in both simulations the same segmented scenario, changing only the radio material assigned to the mesh faces corresponding to the windows, i.e., setting ITU-concrete in one scenario and ITU-glass in the second one.

\section{Comparing ray-based channel simulations}\label{sec:method}

In this section, we propose two metrics to compare two different ray tracing channel simulations between a transmitter and a receiver. As, in general, the cardinality of the set of paths for the compared simulations is not the same, defining a distance metric among them poses different requirements with respect to the direct comparison of fixed-size matrices of parameters. Inspired by the point clouds modeling and processing field, we interpret the ray tracing simulation parameters associated to each propagation path as point clouds embedded in a linear space of suitable dimension.

\subsection{Ray tracing parameters representation}

We denote with $X = \{\mathbf{v}_1, \mathbf{v}_2, \dots, \mathbf{v}_N\}$ and $Y = \{\mathbf{w}_1, \mathbf{w}_2, \dots, \mathbf{w}_M\}$ two sets of ray tracing parameters produced by ray tracing simulations between the same transmitter and receiver over possibly different scenario 3D meshes, where $N$ and $M$ are the total number of propagation paths in the simulation sets $X$ and $Y$, and $\mathbf{v}_i \in X$, $i \in \{1, \dots, N\}$ is a 6-tuple
\begin{equation}\label{eq:rt_params_tuple}
    (P_i, \tau_i, \theta_i^\text{DoD}, \phi_i^\text{DoD}, \theta_i^\text{DoA}, \phi_i^\text{DoA}) \in \mathbb{R}^6,
\end{equation}
where $P_i$ denotes the received power (in dBm), $\tau_i$ is the path delay (in sec.), $(\theta_i^{DoD}, \phi_i^{DoD})$ are the direction of departure (DoD) azimuth and elevation angles, respectively, and $(\theta_i^{DoA}, \phi_i^{DoA})$ are the direction of arrival (DoA) azimuth and elevation angles, respectively.
The components of $Y$ are similarly defined for the second set of ray tracing parameters. As a result, $X$ and $Y$ are subsets of $\mathbb{R}^6$, with possibly different cardinality depending on the number of paths featuring radio propagation between the Tx and the Rx in the first and in the second scenario. The tuple in \eqref{eq:rt_params_tuple} can be purposefully extended to include other simulated quantities like the phase and the Doppler shift. The phase information can be useful to determine the impact in the assignment of different radio materials' properties to the meshes, while the comparison of the Doppler shifts can highlight relevant differences in the presence of diverse motion patterns among the meshes within dynamic environments. These parameters have not been considered in this work, which focuses instead on the analysis of the differences in the power, delay and angular channel profiles.

\subsection{Comparing rays parameters}

We consider $X$ and $Y$ as sets of points in a 6-dimensional linear space. The aim is to determine a distance between $X$ and $Y$ assuming that they generally have different cardinality, and aiming for two relevant properties:
\begin{enumerate}
    \item The distance may be interpretable in terms of the spatial (DoA/DoD azimuth and elevation), temporal (delay) and power features so that their differences can be explicitly analyzed.
    \item The parameters dimensions should be uniformly weighed, or it should be possible to separately assign a custom weight to each feature according to the purpose of the comparison (i.e., the covariance between different dimensions of the point clouds should be normalized to uniformly weight them and should not be arbitrarily elongated depending on the default magnitude of the corresponding quantity).
\end{enumerate}

As such a distance may depend on the relative magnitude of the temporal, spatial and power quantities represented on the dimensions of the 6-dimensional space, we propose to standardize the temporal and power dimensions across the values in the dataset of simulated parameters to compare. Referring to \eqref{eq:rt_params_tuple}, the resulting pre-processed tuple becomes
\begin{equation}
    (\bar{P}_i, \bar{\tau}_i, \theta_i^\text{DoD}, \phi_i^\text{DoD}, \theta_i^\text{DoA}, \phi_i^\text{DoA}),
\end{equation}
where
\begin{equation*}\label{eq:rt_params_tuple_standardized}
    \bar{P}_i = \frac{P_i - \mu_P}{\sigma_P},\qquad \bar{\tau}_i = \frac{\tau_i - \mu_\tau}{\sigma_\tau},
\end{equation*}
and $\mu_P$, $\mu_\tau$, $\sigma_P$, $\sigma_\tau$ are, respectively, the means and the standard deviations of the power and temporal components computed on $X$, or similarly on $Y$.

Considering $\boldv \in X$ and $\boldw \in Y$, we define the following set of distances across the dimensions of \eqref{eq:rt_params_tuple_standardized}:
\begin{equation}\label{eq:features_metrics}
    \begin{gathered}
        d_\tau(\boldv, \boldw) = |\bar{\tau}_\boldv - \bar{\tau}_\boldw|,\\
        d_P(\boldv, \boldw) = |\bar{P}_\boldv - \bar{P}_\boldw|,\\
        d_\text{DoD} = 1 - \hat{\mathbf{u}}(\theta_\boldv^\text{DoD}, \phi_\boldv^\text{DoD}) \cdot \hat{\mathbf{u}}(\theta_\boldw^\text{DoD}, \phi_\boldw^\text{DoD}),\\
        d_\text{DoA} = 1 - \hat{\mathbf{u}}(\theta_\boldv^\text{DoA}, \phi_\boldv^\text{DoA}) \cdot \hat{\mathbf{u}}(\theta_\boldw^\text{DoA}, \phi_\boldw^\text{DoA}),
    \end{gathered}
\end{equation}
where $\hat{\mathbf{u}}(\theta, \phi) = [\cos \phi \cos \theta, \cos \phi \sin \theta, \sin \phi]^\intercal$ is a free space unit vector pointing in the direction identified by azimuth angle $\theta$ and elevation angle $\phi$, the subscripts $\boldv$ and $\boldw$ indicate the features within the respective tuples, $d_\text{DoD}$ is the cosine distance between the free space vectors identified by the two DoD azimuth/elevation angles, and $d_\text{DoA}$ is defined similarly to $d_\text{DoD}$, replacing the DoD azimuth and elevation angles with the corresponding DoA ones.

We keep separate distances across the power, temporal and spatial dimensions to allow for further analyses over separate features, while we aggregate them into a single distance to determine a quantitative overall dissimilarity between two parameters tuples $x \in X$ and $y \in Y$.

Therefore, we define the global metric $d_R: \mathbb{R}^6\times\mathbb{R}^6 \to \mathbb{R} $ between two 6-tuples $\boldv \in X$ and $\boldw \in Y$ as the sum of 4 components related, respectively, to the their temporal, power, and angular (i.e., DoD and DoA) features:
\begin{equation}\label{eq:ray_params_distance}
    d_R(\boldv, \boldw) = d_{\tau}(\boldv, \boldw) + d_P(\boldv, \boldw) + d_{\text{DoD}}(\boldv, \boldw) + d_{\text{DoA}}(\boldv, \boldw),
\end{equation}
where the metrics $d_{\tau}$, $d_P$, $d_{\text{DoD}}$ and $d_{\text{DoA}}$ are defined as in \eqref{eq:features_metrics}. Depending on the temporal and power values distributions, the $d_{\text{DoD}}$ and $d_{\text{DoA}}$ may be scaled in the sum defined in \eqref{eq:ray_params_distance} to further compensate for the metrics range over the standardized temporal and power components.
We notice that, depending on the specific analysis context, the sum components in \eqref{eq:ray_params_distance} can be suitably weighted through a set of hyperparameters to emphasize the importance of some components with respect to the others. In this work, we consider a uniform weighting of the delay, power and angular distance components to assign them the same importance in the computation of the aggregated distance measure $d_R$.

\subsection{Ray tracing simulations distance}

Leveraging the point cloud distance measures in the literature, we propose to compare $X$ and $Y$ using the Hausdorff and the Chamfer distances as they allow to compare sets of possibly different cardinality and they provide summary statistics representing, respectively, the maximum and the average distance between the two sets.

The Hausdorff distance between the finite sets $X$ and $Y$ is
\begin{equation}\label{eq:hausdorff_distance}
    d_H(X, Y) = \frac{1}{2} \max_{\boldv \in X} d(\boldv, \text{NN}(\boldv, Y)) + \frac{1}{2} \max_{\boldw \in Y} d(\boldw, \text{NN}(\boldw, X)),
\end{equation}
where $d(\boldv, \boldw)$ is a distance function between $\boldv \in X$ and $\boldw \in Y$, and
\begin{equation}\label{eq:nn_function}
    \text{NN}(\boldv, Y) = \argmin_{\boldw \in Y} d(\boldv, \boldw)
\end{equation}
is the minimum distance between $\boldv \in X$ and any element of the set $Y$.

The Chamfer distance is similarly defined as
\begin{equation}\label{eq:chamfer_distance}
    d_C(X, Y) = \frac{1}{2 N} \sum_{i = 1}^N d(\boldv_i, \text{NN}(\boldv_i, Y)) + \frac{1}{2 M} \sum_{j = 1}^M d(\boldw_j, \text{NN}(\boldw_j, X)).
\end{equation}

We notice that the Chamfer distance replaces the maximum operation of the Hausdorff distance with the average, keeping, besides this change, the same structure. We consider the bidirectional versions of these distances as they uniformly weigh the distance of the first set of points with respect to the second one and vice versa.

We denote our specific distance measures \eqref{eq:hausdorff_distance} and \eqref{eq:chamfer_distance} in the ray tracing simulations setting as \textbf{Hausdorff-RT} (HRT) and \textbf{Chamfer-RT} (CRT), respectively, to distinguish them from the definitions operating on generic sets. For both, we set the distance $d$ equal to $d_R$, where the latter is defined in \eqref{eq:ray_params_distance}, and the elements over which the two metrics are computed are the 6-dimensional vectors representing rays parameters defined in \eqref{eq:rt_params_tuple_standardized}.

Using the proposed metric $d_R$, the HRT and CRT distances allow to keep information on both the global distances resulting from the use of $d_R$ and of distances $d_{\tau}$, $d_P$, $d_{\text{DoD}}$ and $d_{\text{DoA}}$ composing $d_R$ and providing quantitative details on the power, temporal and spatial dissimilarity between two sets of rays parameterized as in \eqref{eq:rt_params_tuple_standardized}. We remark that we used the composite distance $d_R$ for the computation of the HRT and CRT distances, and, in particular, to evaluate the nearest neighbor functions for the two methods, while we kept track of the separate components resulting from those nearest neighbor assignments in the two metrics. This allows to determine the nearest neighbors by jointly taking into account the power, temporal and spatial ray features, while storing the resulting separate features' metrics for further analyses and plotting purposes. If only one of the distance components in \eqref{eq:ray_params_distance} is used to determine the nearest neighbors in \eqref{eq:hausdorff_distance} or \eqref{eq:chamfer_distance} instead of the aggregated metric $d_R$, the assignment takes only into account a specific ray tracing channel feature. We notice that this would result in a different type of assessment that can be useful as alternative comparison type when the focus of the analysis in on a single feature. In Section \ref{sec:results}, we refer to these distances as HRT or CRT distances over power, delay, DoD or DoA, depending on the separate features over which the metric has been computed.

In Section \ref{sec:results}, we compute the HRT and CRT distances to assess the ray tracing differences deriving from two kinds of environmental changes in the considered scenario, i.e., the addition of parked vehicles to the basic setting composed only by buildings and ground, and the segmentation of the buildings facades to assign different materials to the windows meshes.

\section{Simulation results}\label{sec:results}

In this section, we propose numerical results based on the HRT and CRT distance metrics between ray tracing simulations introduced in Section \ref{sec:method} accounting for power, temporal and angular features in the simulated rays. First, we detail the simulation setup used to produce the ray tracing simulations. Then, we consider two types of environmental changes in the base propagation scenario described in Section \ref{sec:scenario}: (i) the addition of a set of parked vehicle meshes to the environment, detailed in Section \ref{sec:parked_vehicles_modeling}, and (ii) the segmentation of the buildings facades to separate the windows from the rest of the building, in order to assign a different radio-material to the windows meshes, discussed in Section \ref{sec:windows_segmentation}. In both cases, we assess the radio propagation differences over ray tracing simulations considering a fixed transmitter base station located over the roof of a building. We focus on two receivers configurations: a dense grid of receivers positioned at fixed height spanning the whole scenario, and receivers located in common vehicular positions derived using a vehicular traffic simulator, separately discussing the two conditions.

\subsection{Simulation setup}\label{sec:simulation_setup}
We describe in the following the steps defining the simulation setup considered in this work. We first detail the workflow adopted to achieve vehicular traffic simulations. Then, we focus on the assignment of radio materials to the scenarios' meshes and to the simulation of the ray tracing parameters needed for the comparative analyses proposed in Sections \ref{sec:results_parked} and \ref{sec:results_windows}.

\begin{figure}
    \centering
    \includegraphics[width=\linewidth]{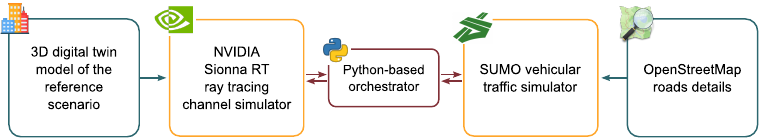}
    \caption{Simulation workflow.}
    \label{fig:simulation_pipeline}
\end{figure}

\textbf{Vehicular traffic simulations}: To pursue realistic vehicular traffic simulations, we used the simulation of urban mobility (SUMO) \cite{SUMO2018} vehicular traffic simulator. We retrieved the traffic network topology of the considered area of Milan, Italy from OpenStreetMap \cite{openstreetmap} (OSM). After converting the OSM data into a SUMO traffic network, we employed the Python interface of SUMO to set up the traffic simulations and to collect the simulated data. For each simulation time step, SUMO provided information for a set of simulated vehicles, among which we gathered their position, which we used to set the positions of the receivers in Sionna RT to define one of the types of considered ray tracing simulations as detailed in the following.

\textbf{Radio materials assignment}: As different mesh types are involved in our experiments, we consider the most suitable radio materials among the ones specified by ITU in the recommendation \cite{ITU-R-2040}. We considered the ITU concrete radio material for the buildings and the ground, ITU glass for buildings windows and vehicles windows, and perfect electric conductor (PEC) for the remaining vehicle components. We notice that the radio material parameters may depend on the adopted carrier frequency and that, in the simulations carried out in this work, we retrieved the radio material parameters at 28 GHz, used as simulation carrier frequency in this work.

\textbf{Ray tracing channel parameters simulation}: We used the NVIDIA Sionna RT \cite{hoydis2023sionna} ray tracing simulator to achieve accurate wireless propagation simulations while retrieving the rays parameters required for our comparisons. In our experiments, we employed Sionna RT v0.18.0, which accounts for specular reflection, diffraction and diffuse scattering interaction types with the environment. Sionna RT relies on the Mitsuba 3 \cite{mitsuba} differentiable rendering system and on the TensorFlow \cite{tensorflow} open source machine learning library. The differentiability of the ray tracing procedures implemented in Sionna RT allow its straightforward use within end-to-end trainable deep learning pipelines, directly enabled also by its TensorFlow-based implementation. As a result, Sionna RT supports GPU acceleration to speed-up ray tracing simulations.

Two methods are currently supported by Sionna RT to provide the set of initial candidate rays: (i) an exhaustive mode, evaluating every possible combination of mesh primitives, and (ii) Fibonacci, based on shooting and bouncing rays (SBR) \cite{hoydis2023sionna} to efficiently list a set of candidate paths but providing no guarantees that every possible path is found. For the efficiency of the ray tracing experiments, we employ the Fibonacci method, which is also more scalable to scenarios of wider dimensions, where the exhaustive option would be intractable. We notice that Sionna RT's Fibonacci method is based on the deterministic sampling of a sphere by means of a Fibonacci lattice to determine the initial directions of the rays in the SBR procedure.

We preprocessed the scenarios' meshes using the Blender \cite{blender} 3D modeling software to specify the meshes radio materials and to export them in the Mitsuba XML file format. We then imported the scenario in Sionna RT setting the ray tracing simulation parameters and the Tx and Rx antennas locations for the two considered simulation types: for \emph{grid-based} simulations, we defined a grid of receivers at fixed height, while, for the \emph{vehicular} ones, we run step-by-step simulations positioning the receivers at the vehicular positions provided by the vehicular traffic simulator at each discrete simulation time step. The grid resolution is $2\times 2$ m and the grid is set at a height of 1.5 m. The grid points that resulted within the buildings or vehicles meshes were filtered out before simulation. For both the simulation conditions, we considered a single Tx base station located as specified in Section \ref{sec:scenario}. The simulation parameters are reported in Table \ref{tab:simulation_parameters}.

\begin{table}[t!]
\centering
\caption{Ray tracing simulation parameters. For the interaction types, R stands for reflection, D for diffraction and DS for diffuse scattering.}
\begin{tabular}{r c}
\noalign{\smallskip}
\textbf{Simulation parameter} & \textbf{Value}\\
\hline
\noalign{\smallskip}
Carrier frequency & 28 GHz\\
BS height from the ground & 21.7 m\\
Rx height from the ground & 1.5 m\\
Rx grid resolution & $2\times 2$ m\\
SUMO simulation sampling time & 0.1 s\\
Number of initial sampled rays & $10^6$\\
Sionna RT ray tracing method & Fibonacci\\
Selected interation types & R, D, DS
\end{tabular}

\label{tab:simulation_parameters}
\end{table}

\subsection{Assessing radio differences in parked vehicles modeling}\label{sec:results_parked}

We provide here the results obtained through the computation of the HRT and CRT metrics on grid-based and vehicular simulations to compare the base reference scenario with the one enriched by a set of parked vehicles as described in Section \ref{sec:parked_vehicles_modeling}.

\begin{figure*}[th]
    \centering
    \subfloat[][Delay]{\includegraphics[width=.49\textwidth]{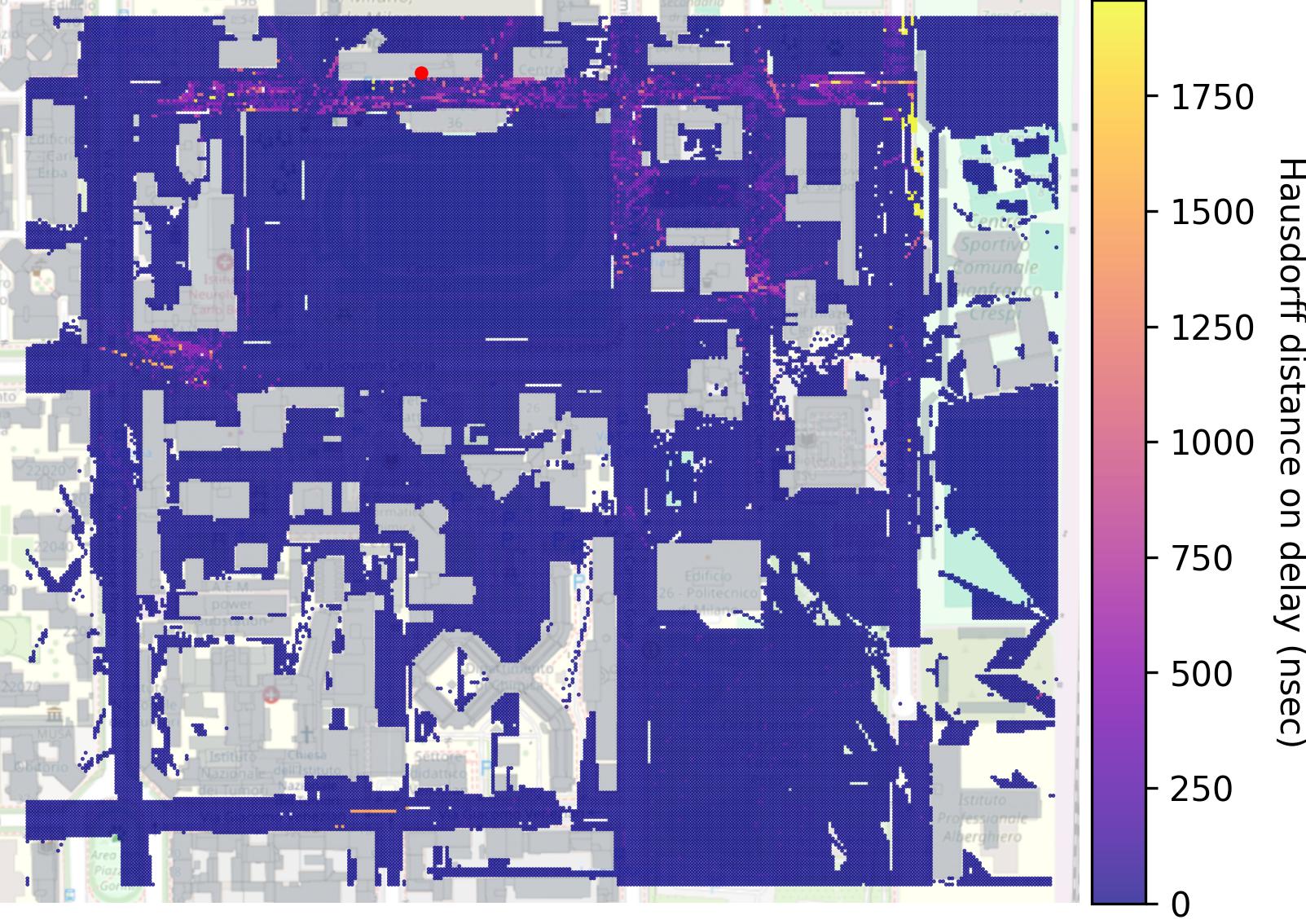}}
    \hspace{0.1cm}
    \subfloat[][Power (dB)]{\includegraphics[width=.47\textwidth]{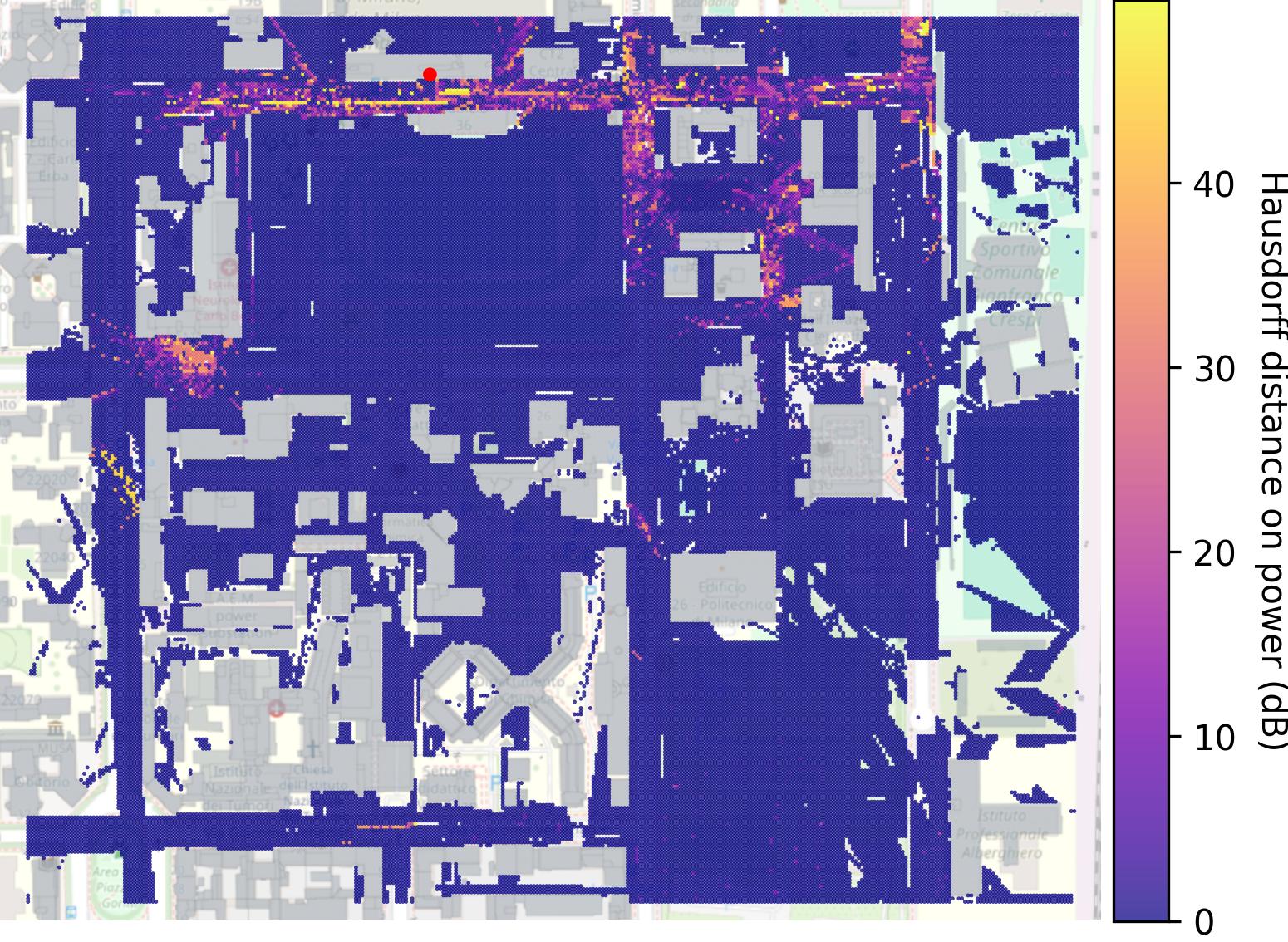}}\\
    \subfloat[][Direction of Departure (deg)]{\includegraphics[width=.49\textwidth]{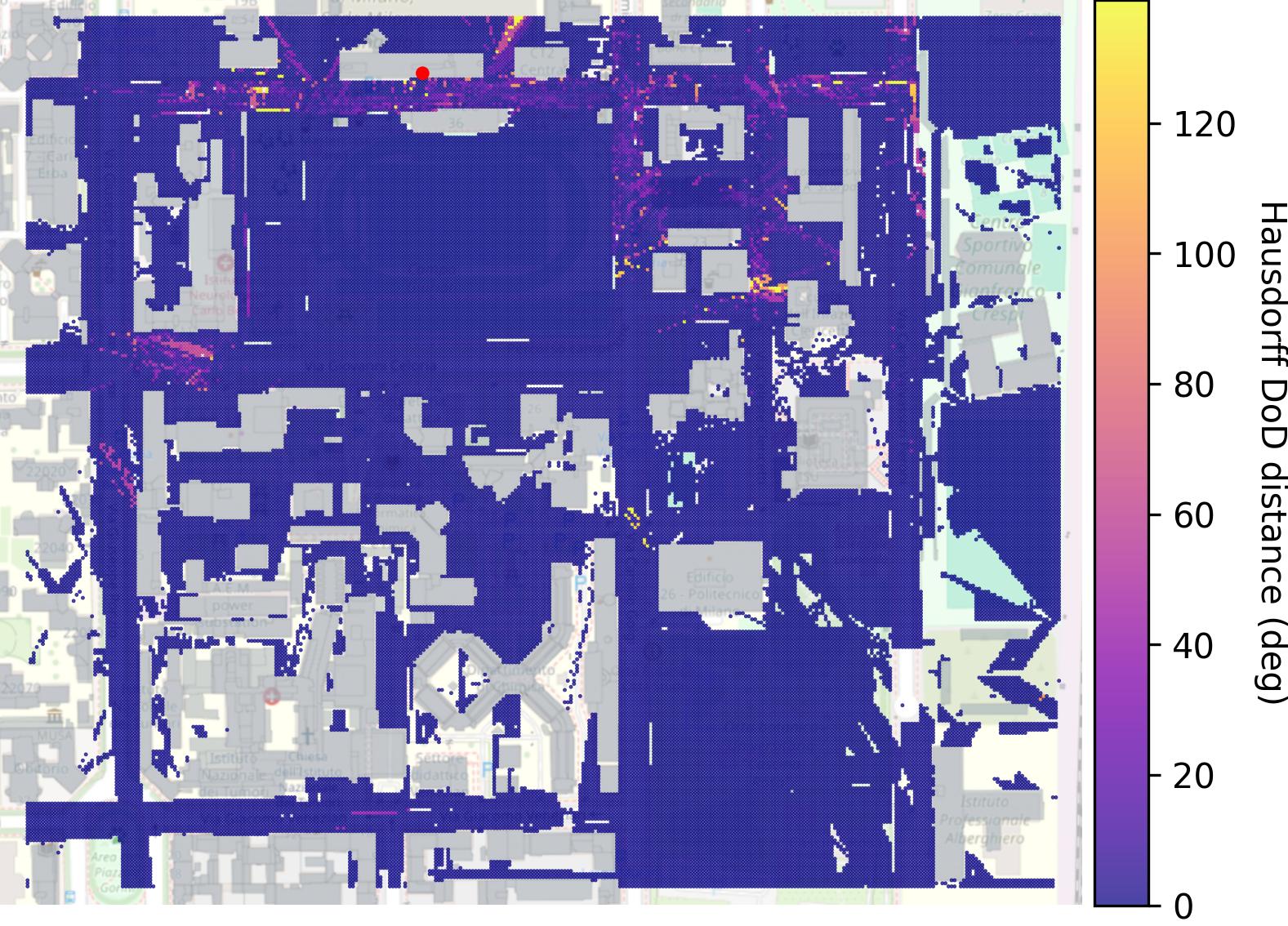}}
    \hspace{0.1cm}
    \subfloat[][Direction of Arrival (deg)]{\includegraphics[width=.49\textwidth]{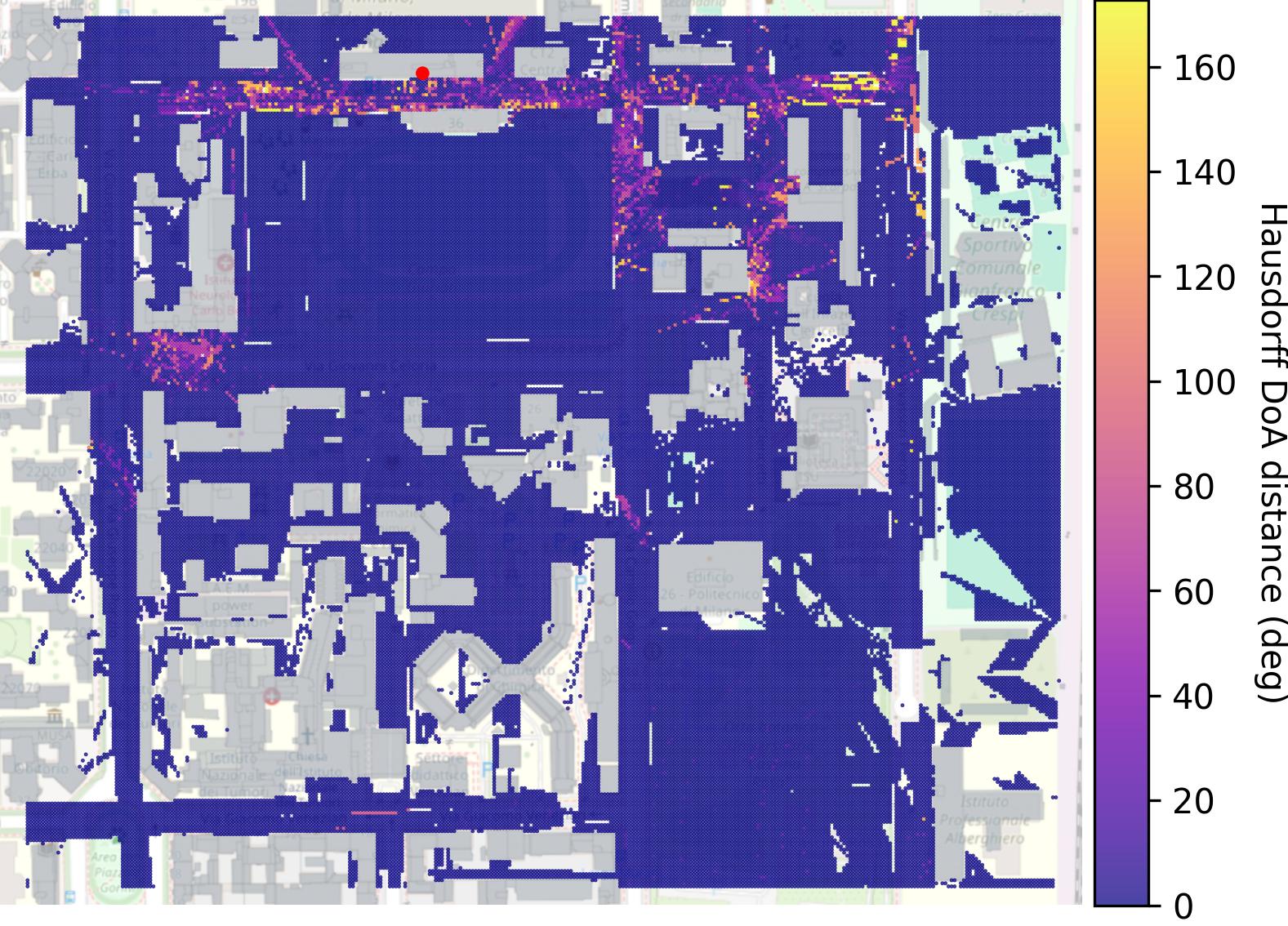}}\\
    \caption{Evaluated Hausdorff distance (HRT) over the delay, power and angular features for the grid-based simulation on the scenario enriched with parked vehicles. The red point indicates the position of the BS, while the colored areas represent the distance metric, whose range of values is indicated by the color bars.}
    \label{fig:hausdorff_parked_grid_results}
\end{figure*}

\begin{figure*}[th]
    \centering
    \subfloat[][Delay]{\includegraphics[width=.49\textwidth]{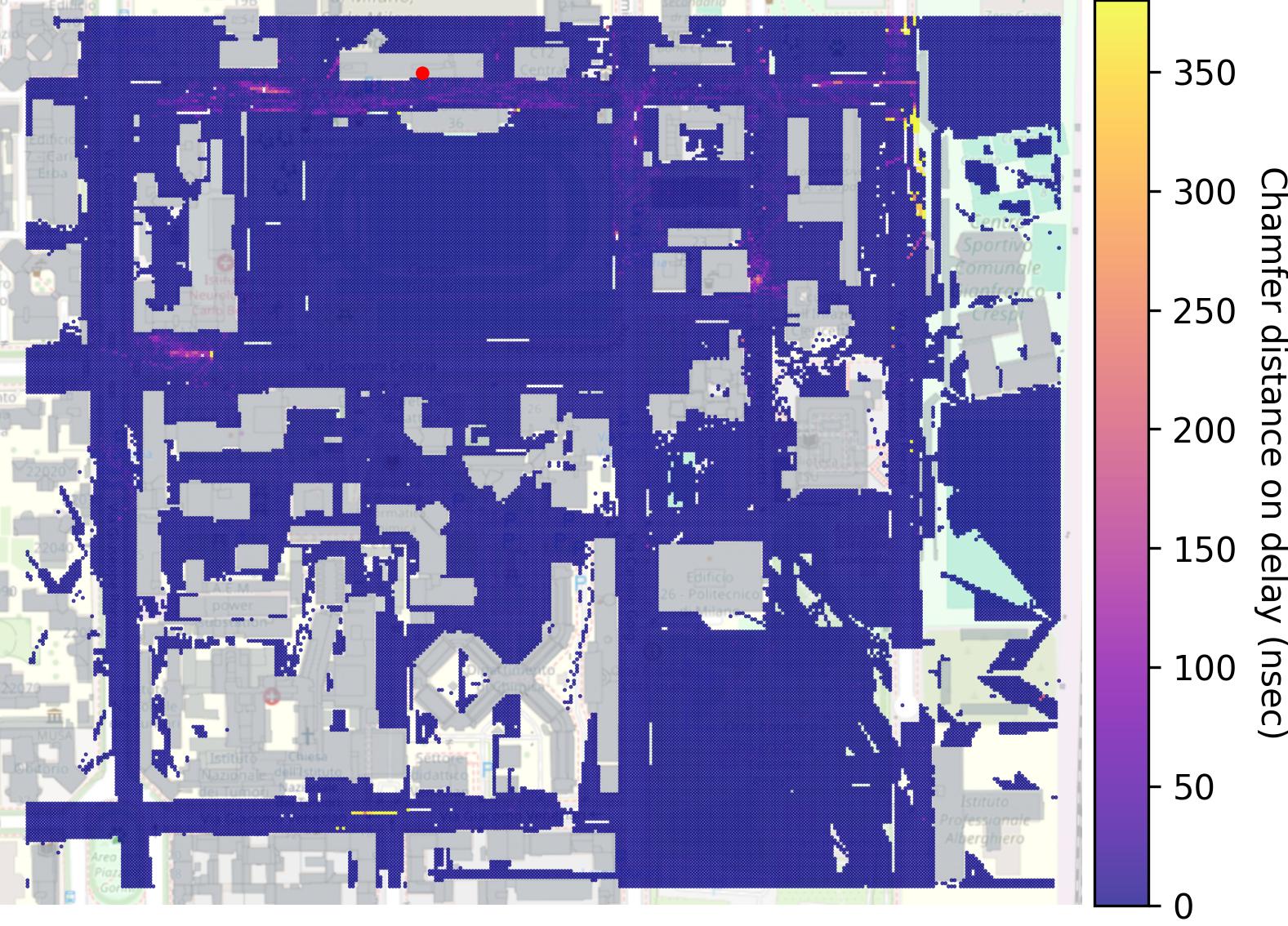}}
    \hspace{0.1cm}
    \subfloat[][Power (dB)]{\includegraphics[width=.47\textwidth]{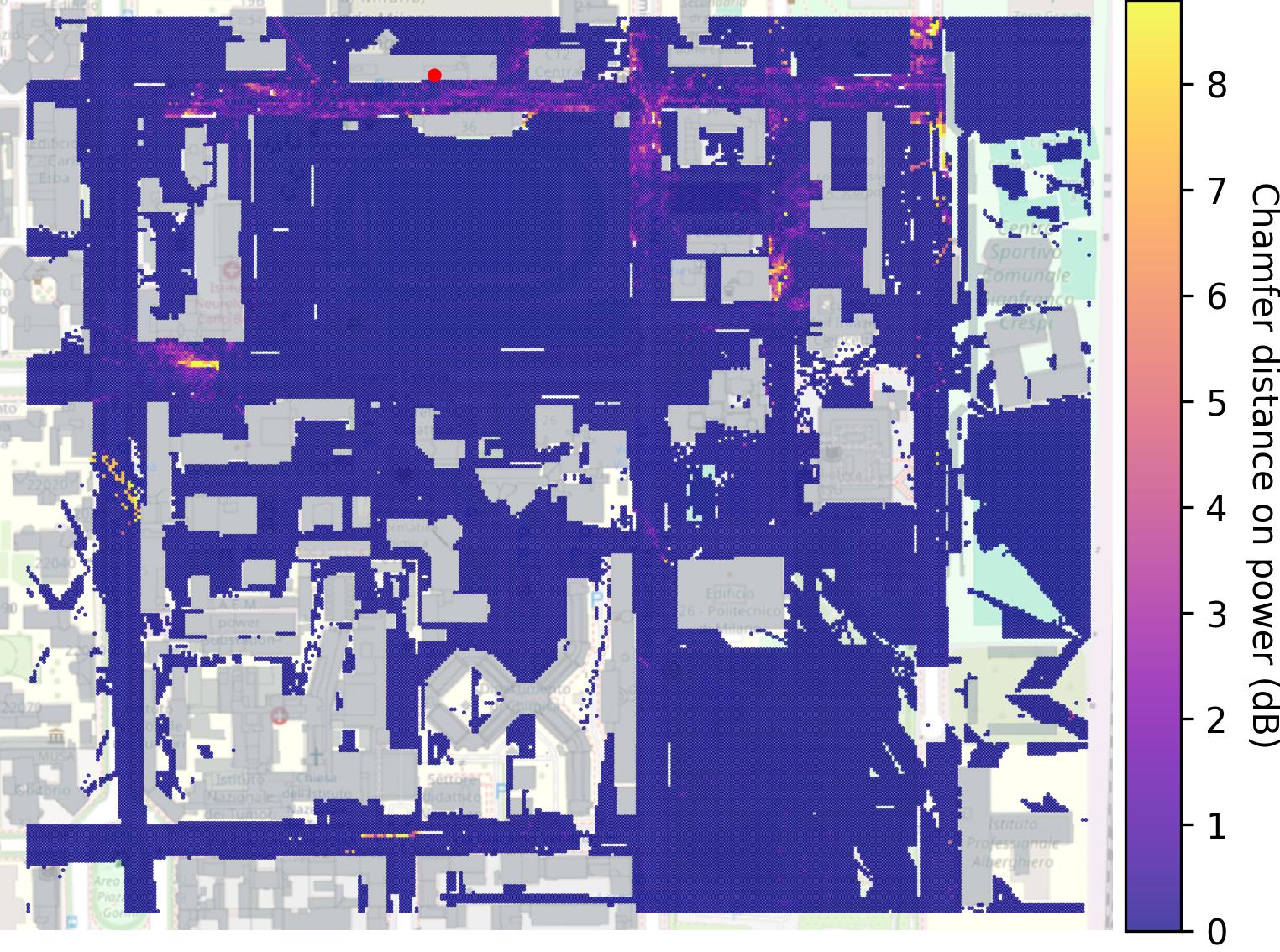}}\\
    \subfloat[][Direction of Departure (deg)]{\includegraphics[width=.49\textwidth]{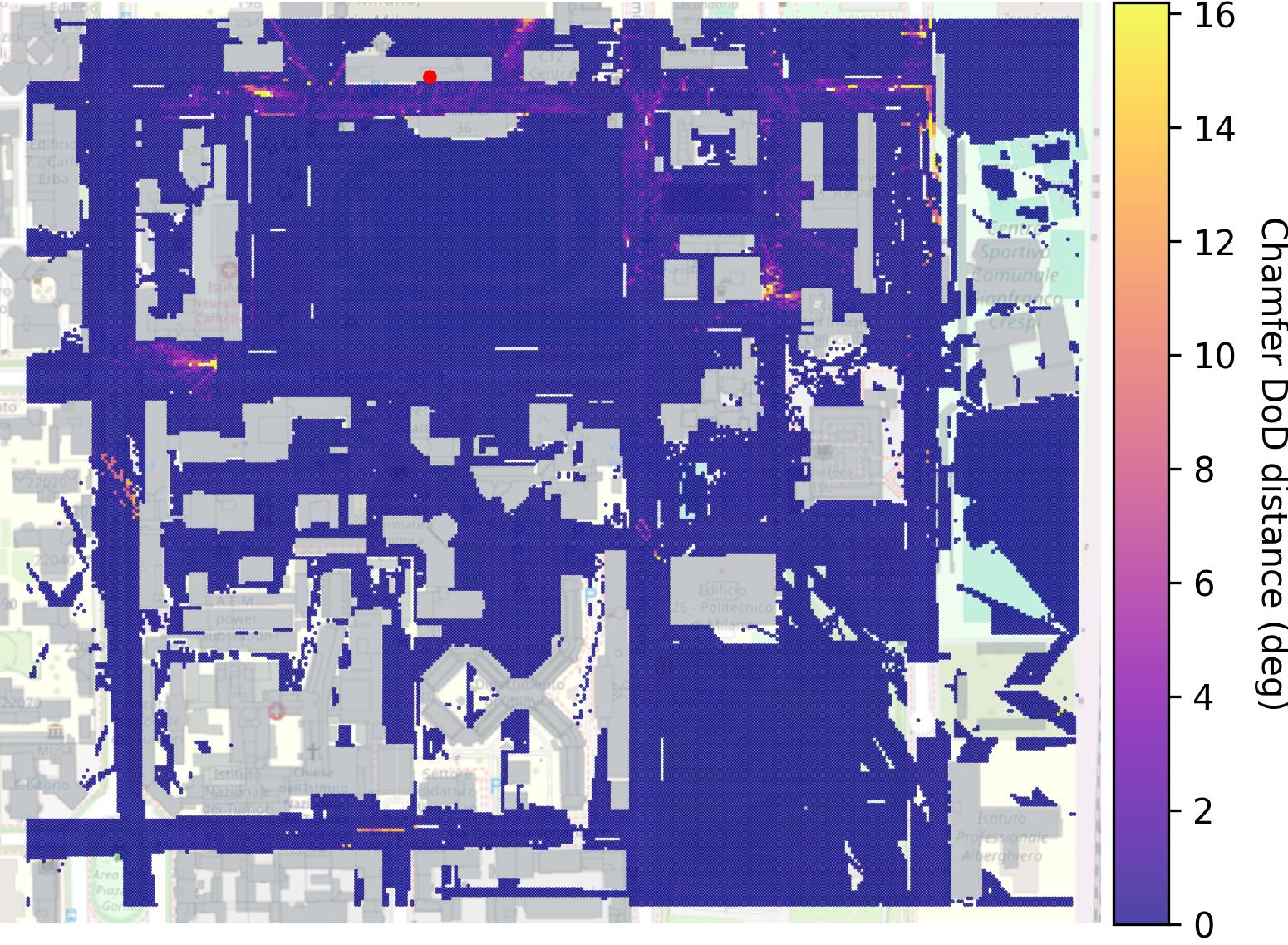}}
    \hspace{0.1cm}
    \subfloat[][Direction of Arrival (deg)]{\includegraphics[width=.49\textwidth]{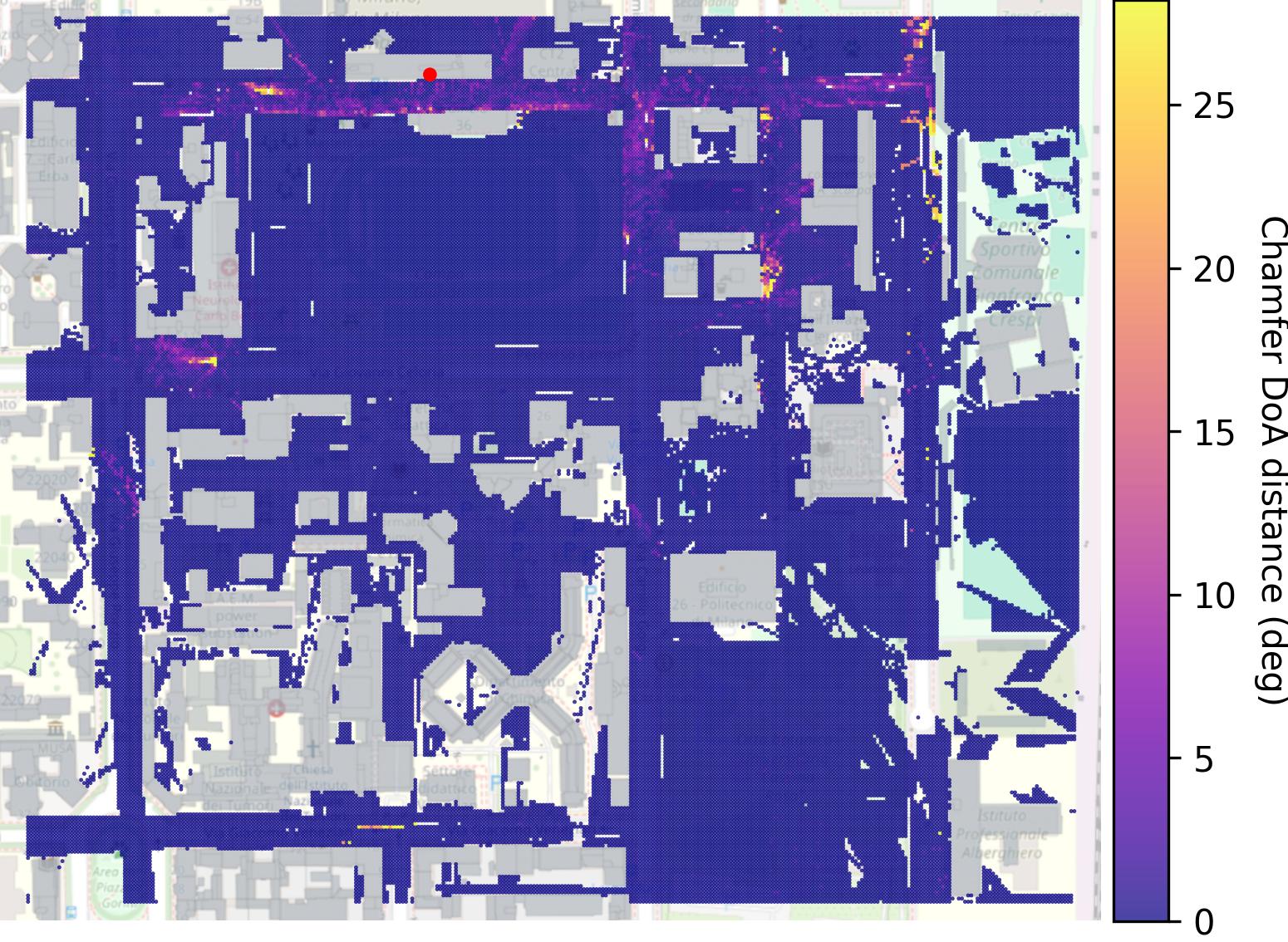}}\\
    \caption{Evaluated Chamfer distance (CRT) over the delay, power and angular features for the grid-based simulation on the scenario enriched with parked vehicles. The red point indicates the position of the BS, while the colored areas represent the distance metric, whose range of values is indicated by the color bars.}
    \label{fig:chamfer_parked_grid_results}
\end{figure*}

\textbf{Grid-based ray tracing simulations}: The comparison results for the grid-based simulations are reported in Fig. \ref{fig:hausdorff_parked_grid_results} and Fig. \ref{fig:chamfer_parked_grid_results}, which depict the HRT and the CRT, respectively, over all the Rx points that resulted in at least one path from the ray tracing simulations. In the proposed figures, the results were not filtered, so that the presence of evaluated points directly corresponds to the radio coverage simulated through NVIDIA Sionna RT over the grid points. As specified in Section \ref{sec:method}, the distances related to the separate temporal, power and angular (DoD and DoA) features are computed during the evaluation of the joint Hausdorff and Chamfer distances using the joint distance $d_R$ as in \eqref{eq:ray_params_distance} to determine the nearest neighbors within the HRT and CRT, and updating the corresponding separate features statistics considering that nearest neighbor assignment.

For both the HRT and CRT, the main differences over delay, power, DoD and DoA are concentrated in the neighborhood of the BS, with few or no relevant differences in the lower part of the scenario. This confirms that the impact of parked vehicles modeling is less relevant farther from the base station. Since, as common for wide scenarios, a SBR method has been used for ray tracing to list the initial candidate path directions (i.e., a fixed number of rays is propagated towards a discrete set of directions at the Tx), this is consistent with the lower probability to hit or reach parked vehicles that are far from the Tx base station, both for the high path loss emerging at high frequencies and owing to the number number of paths that are propagated from the Tx towards the receivers, usually limited to balance simulation efficiency and accuracy. Therefore, in the following, we focus on ray tracing simulation differences in the neighborhood of the BS.

We remark that the CRT represents a mean value for the distance between the features of the paths in the first simulation set (reference scenario) and the ones in the second set (scenario enriched with parked vehicles). The HRT represents instead a maximum value over a similar computational procedure. Therefore, the values of the HRT are always higher with respect to the CRT ones. Nevertheless, we notice that similar patterns arise on the grid for both the HRT and the CRT in Figs.~\ref{fig:hausdorff_parked_grid_results} and \ref{fig:chamfer_parked_grid_results}, respectively. This shows that both the distance measures are consistent in highlighting similar areas in the scenarios over which the ray tracing simulations differ.

The differences in power show an average value of $\approx 16 dB$ in the neighborhood of the BS, where we highlight that the HRT takes into account all the rays in the compared simulations. This implies that new rays may have emerged or may have been dropped by the inclusion of the parked vehicles meshes, causing the increase in the retrieved HRT values. Similar patterns can be observed in the HRT over path delays, with an average value of $\approx 277$~ns near to the BS, and in DoD and DoA, with an average value of $\approx 46$~deg and $\approx 29$~deg, respectively, in the neighborhood of the base station.

Also for CRT, the patterns across power, delay and angular differences is similar on the whole scenario, with higher differences in the neighborhood of the BS. Near to the BS, the observed average value of the CRT power distance is $\approx 4.2$ dB. For the delay, the average value in the same area is $\approx 47$ ns, while for DoD and DoA the CRT distance is $\approx 10.2$ deg and $\approx 11.6$~deg, respectively.

\begin{figure*}[t!]
    \centering
    \subfloat[][Selected vehicular trajectory.]{\includegraphics[width=.49\textwidth]{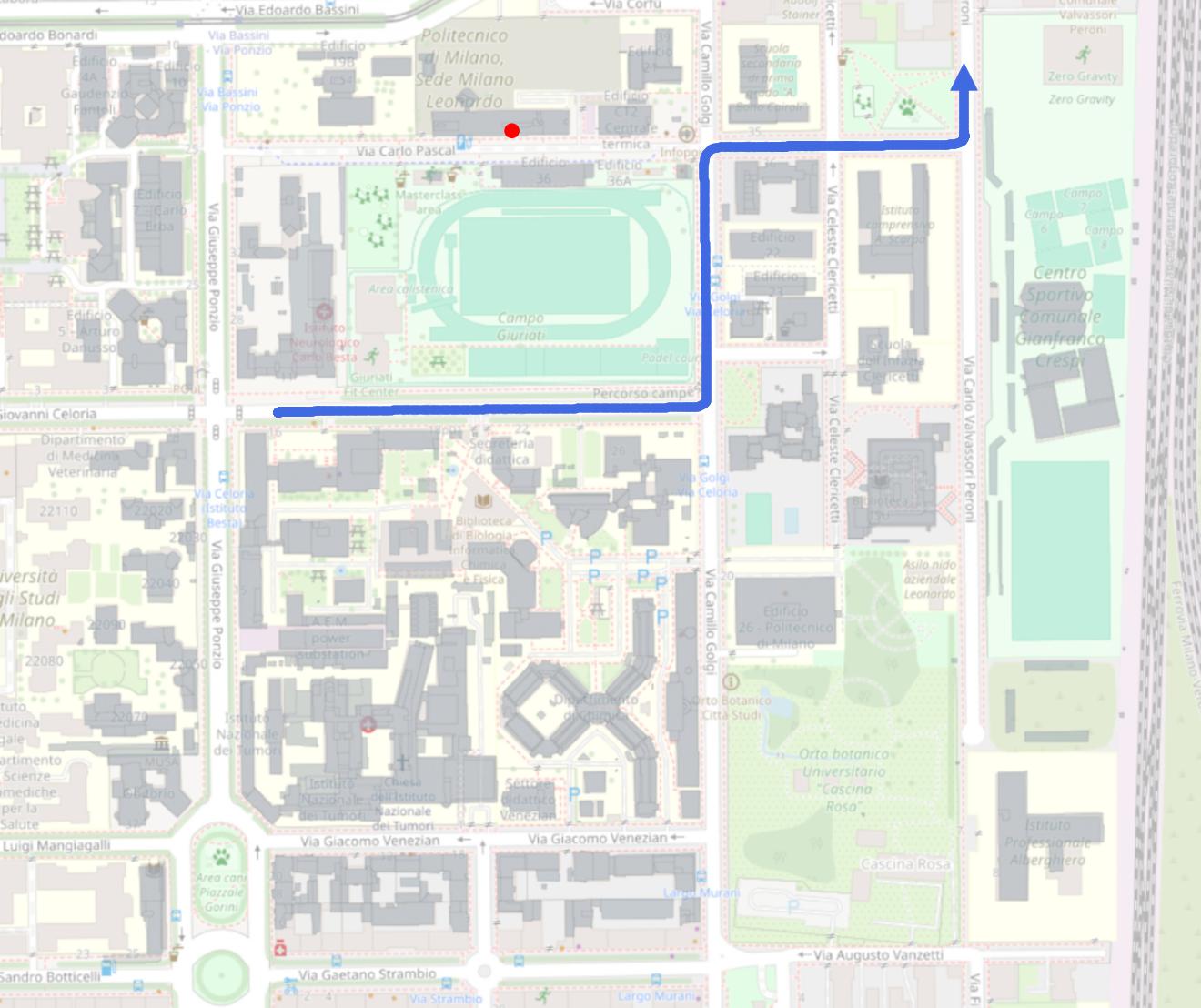}}
    \hspace{0.1cm}
    \subfloat[][Power distance (dB)]{\includegraphics[width=.49\textwidth]{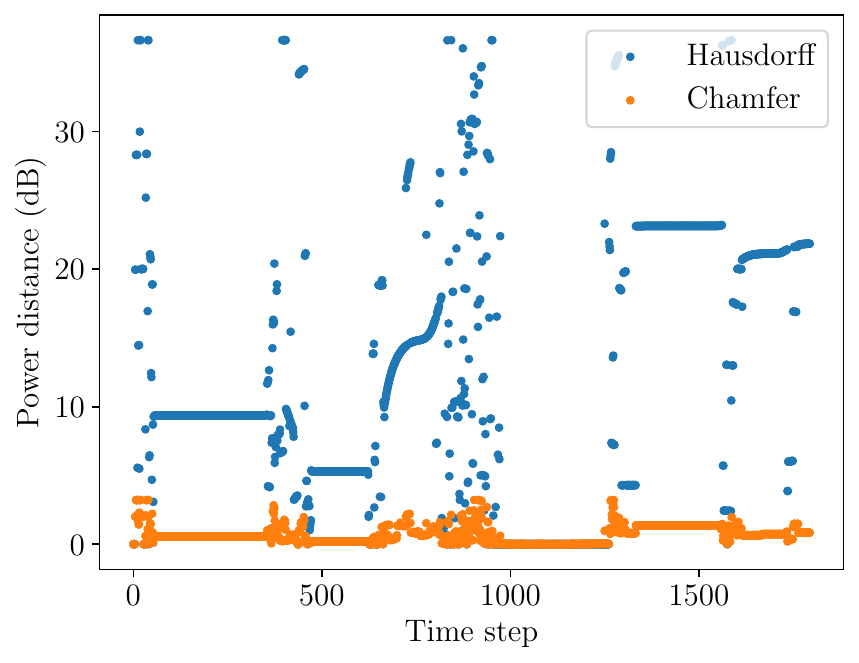}}\\
    \subfloat[][Delay distance (nsec)]{\includegraphics[width=.49\textwidth]{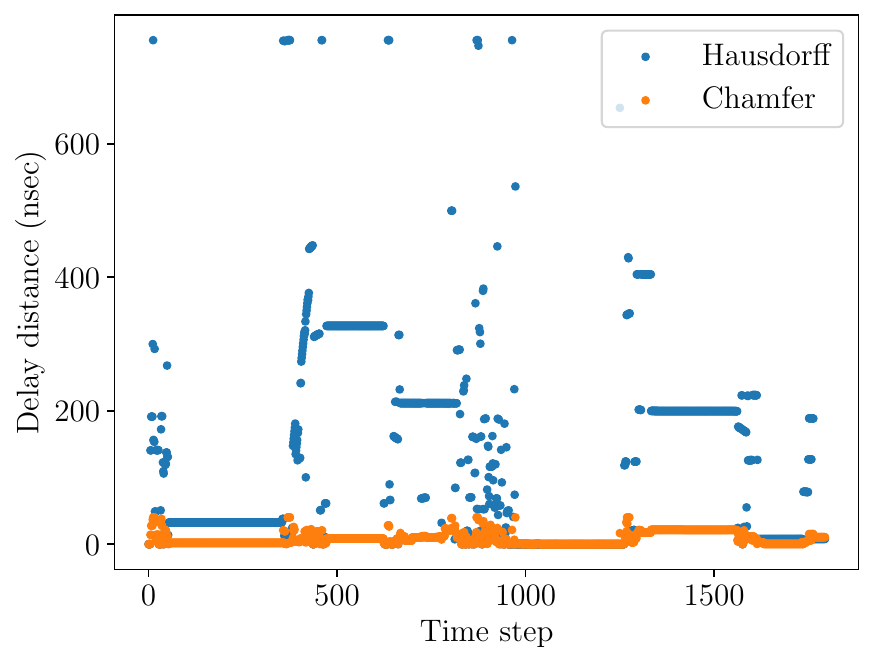}}
    \hspace{0.1cm}
    \subfloat[][DoA distance (deg)]{\includegraphics[width=.49\textwidth]{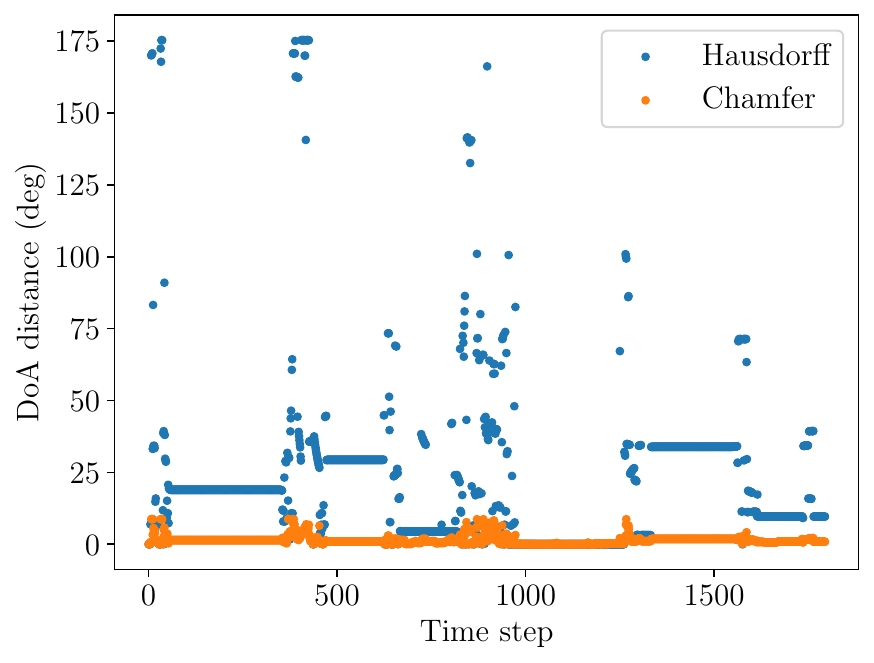}}\\
    \caption{Evaluated Hausdorff (HRT) and Chamfer (CRT) distances over power for a vehicular trajectory on the scenario enriched with parked vehicles. In (a), the red point indicates the position of the BS, the vehicular trajectory is represented in blue, and the triangle on one of the extremes of the trajectory indicates its start.}
    \label{fig:parked_veh_results}
\end{figure*}

\textbf{Vehicular ray tracing simulations}: We provide in Fig. \ref{fig:parked_veh_results} the obtained results on the evaluation of HRT and CRT considering a set of Rxs located over a vehicular trajectory of a sedan crossing the reference scenario. Both the HRT and CRT present a set of patterns owing to the motion of the vehicle on its trajectory. This provides a further information with respect to the grid-based simulations, which instead were aimed at fingerprinting the simulation distances over the whole scenario. Consistently with the grid-based simulations, the HRT distance on power shows a maximum extension of $35$ dB and the CRT is similarly limited depending on the crossed area of the scenario. The HRT and CRT distances on delay and angles show also a similar behavior, with featuring patterns owing to the continuity of the vehicle trajectory. The figure only provides the DoA distances as the DoD HRT and CRT distances presented similar patterns, outlining the same behavior of the corresponding areas in the corresponding results proposed for the grid-based simulation.

On the considered trajectory, we observed an average power HRT distance of $\approx 12$ dB and an average power CRT distance of $\approx 0.70$ dB. On the delay features, the computed mean HRT distance is $\approx 130$ ns, while the corresponding mean CRT distance is $\approx 8.6$ ns. On the DoDs, we observed a mean HRT distance of $\approx 21.7$ deg and a CRT distance of $\approx 0.8$ deg, while on the DoAs, the mean HRT distance is of $\approx 23$ deg and the mean CRT of $\approx 1.4$ deg.

\begin{figure*}[t!]
    \centering
    \subfloat[][Grid-based simulation]{\includegraphics[width=.49\textwidth]{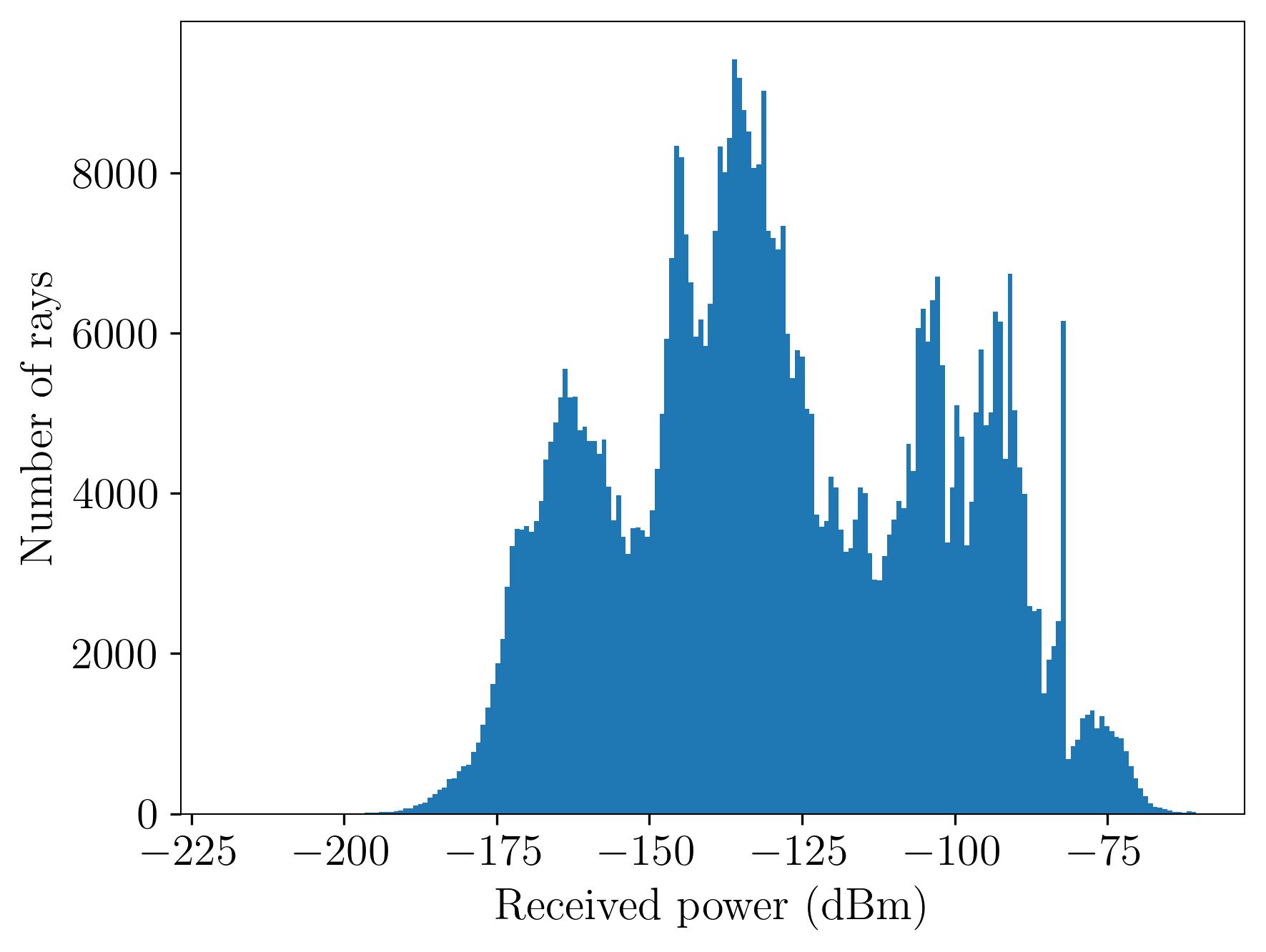}}
    \hspace{0.1cm}
    \subfloat[][Vehicular simulation]{\includegraphics[width=.49\textwidth]{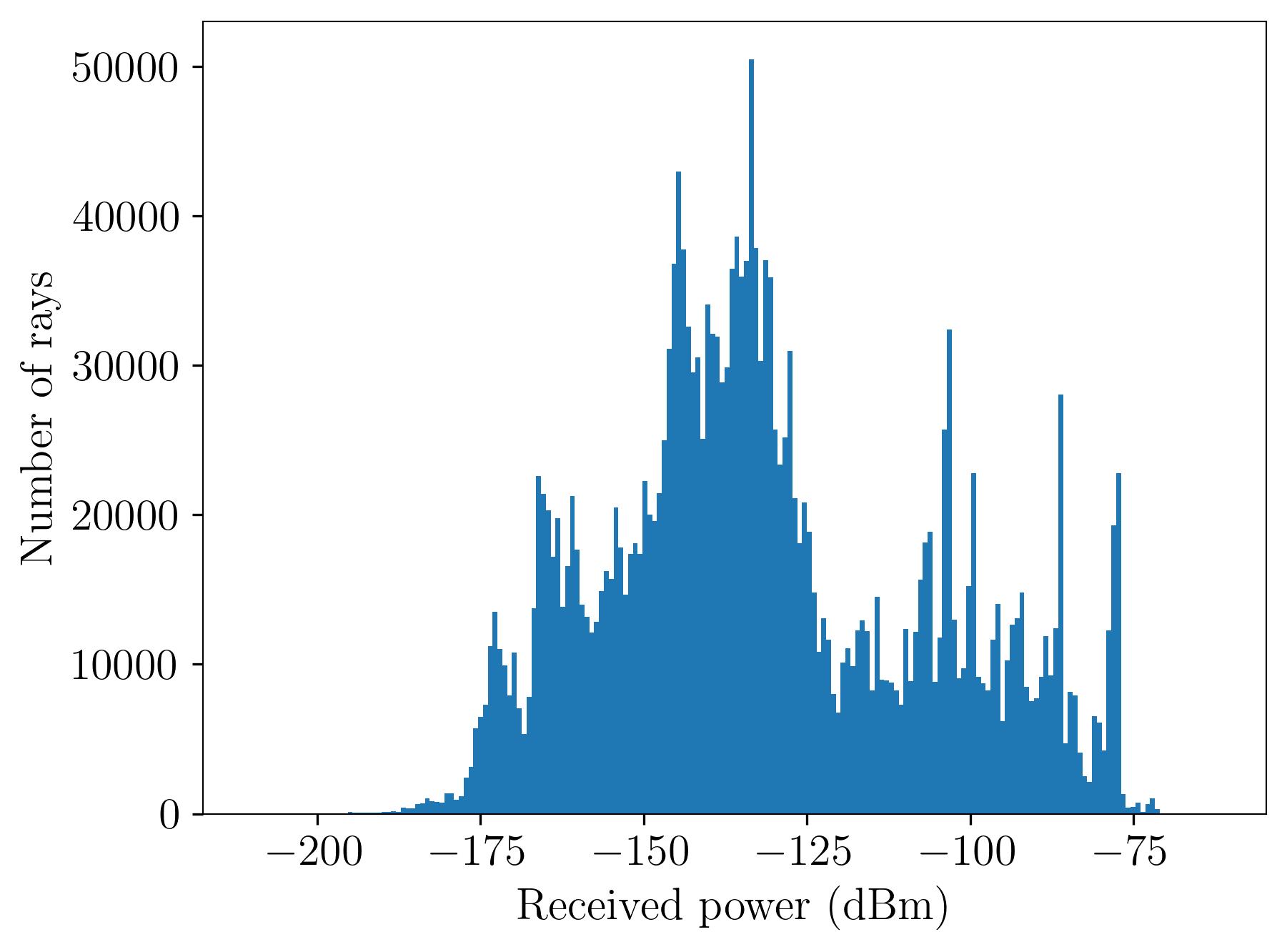}}\\
    \caption{Power profiles of the rays produced by ray tracing simulations using Sionna RT over the scenario with parked vehicles for grid-based (a) and vehicular simulations (b).}
    \label{fig:rays_power_profiles}
\end{figure*}

Simulations have been performed over a wide set of trajectories produced by the SUMO vehicular traffic simulator. In this discussion, we considered one of them as representative of the observed patterns over power, delay, and angular features. We observed similar patterns also for the remaining vehicular trajectories, with higher differences in the neighborhood of the BS and few or no ray tracing differences in the areas farther from the BS, consistently with the grid-based simulations proposed above.

For completeness, we provide in Fig. \ref{fig:rays_power_profiles} the power profiles of the paths deriving from the grid-based and vehicular simulations on the scenario enriched with parked vehicles. Similar patterns have been observed also on the scenario enriched with windows segmentation.

\subsection{Assessing radio differences in windows segmentation modeling}\label{sec:results_windows}

We provide here the results obtained through the computation of the HRT and CRT metrics on grid-based and vehicular simulations to compare the base reference scenario with the one enriched by the buildings windows segmentation as described in Section \ref{sec:windows_segmentation}.

\begin{figure*}[t!]
    \centering
    \subfloat[][Hausdorff distance]{\includegraphics[width=.49\textwidth]{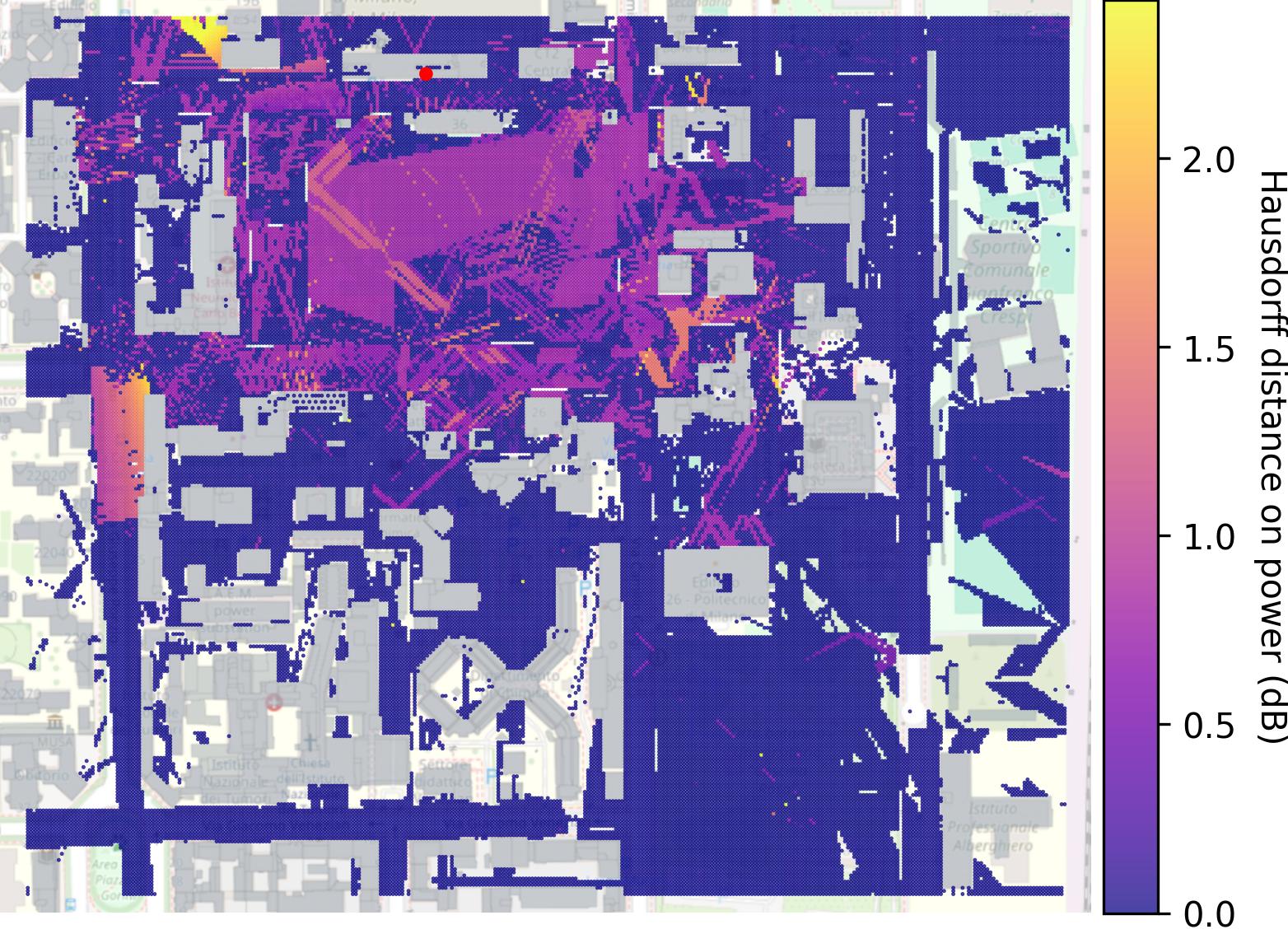}}
    \hspace{0.1cm}
    \subfloat[][Chamfer distance]{\includegraphics[width=.49\textwidth]{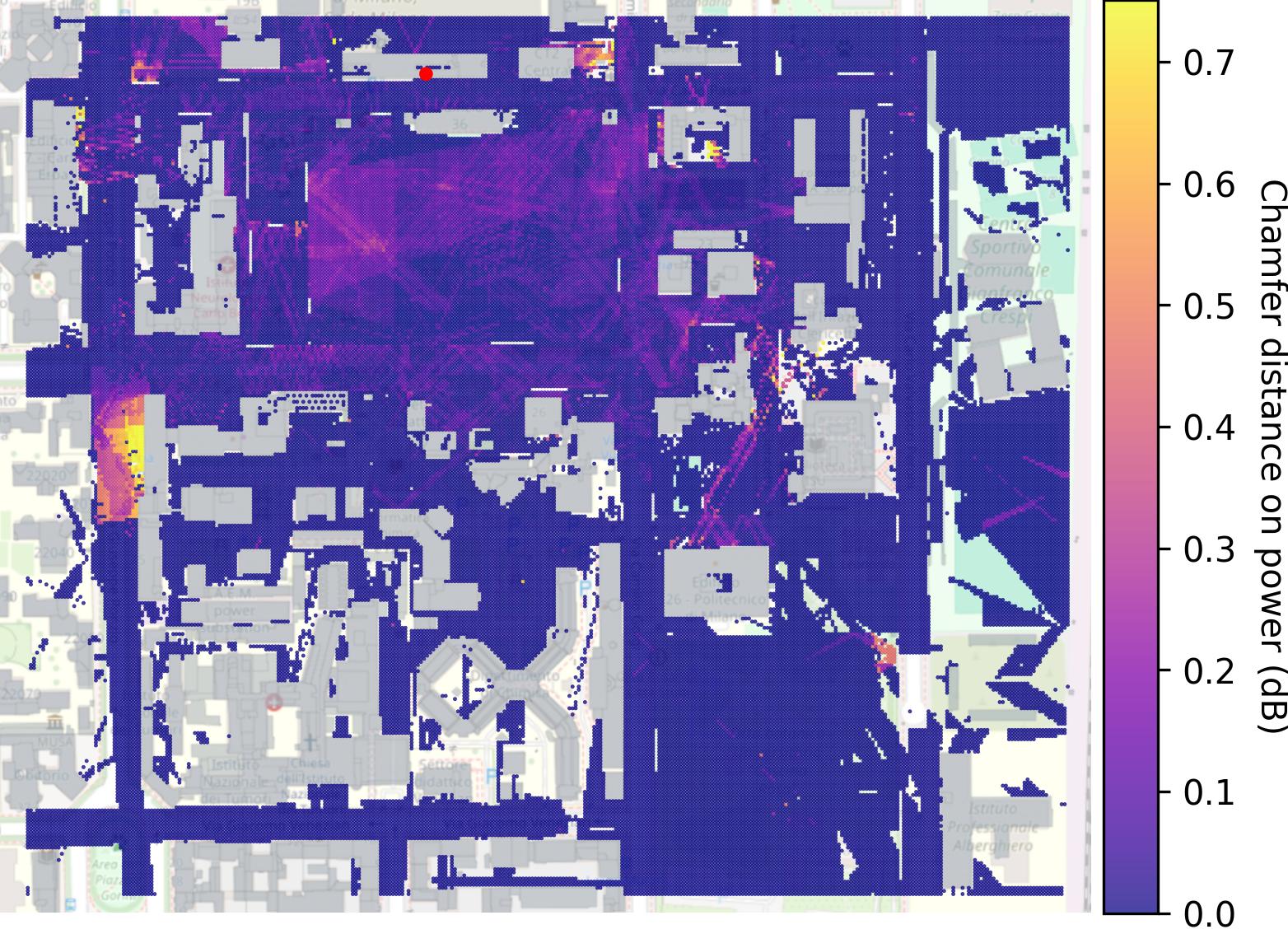}}\\
    \caption{Evaluated Hausdorff (HRT) and Chamfer (CRT) distances over power for the grid-based simulation on the scenario enriched with buildings windows segmentation. The red point indicates the position of the BS, while the colored areas represent the distance metric, whose range of values is indicated by the color bars.}
    \label{fig:windows_grid_results}
\end{figure*}

\textbf{Grid-based ray tracing simulations}: We provide in Fig. \ref{fig:windows_grid_results} the comparison results over grid-based ray tracing simulations between the reference scenario and the scenario comprising segmented windows. Since no differences have been observed on rays delays and angles, we focus the discussion on the power differences between the two scenarios. We notice that the lack of diversity in the temporal and angular features is consistent with the applied environmental changes because, as discussed above, only the radio material assigned to the windows mesh components varies between the two scenarios. Also, the simulated interaction types providing the most relevant interactions with the environment---i.e., specular reflection and diffraction---are not affected on angles and delays by the changes in the material properties.

Also in this case, the majority of ray tracing differences is concentrated near the BS. Nevertheless, compared to the scenario enriched with parked vehicles discussed in Section \ref{sec:results_parked}, the power difference patterns produced by windows segmentation appear different and are distributed on different areas of the environment. This highlights the different impact of an environmental change introduced at different heights, as parked vehicles are near to the ground while windows can be spanned over the whole facade of a building. Moreover, windows have different extensions and mesh structure (mostly planar) with respect to parked vehicles, inducing a diverse variation in the radio propagation patterns.

With respect to the scenario enriched with parked vehicles, the power distance range is more limited, since it is only related to the different radio material set for windows meshes. We remark that for this scenario no meshes have been changed with respect to the base one, so that the only modification is in the assignment of the radio material depending on the windows and non-windows meshes. We observed a maximum HRT power distance of $\approx 2.42$~dB and an average HRT distance of $\approx 0.81$ dB, while the maximum CRT distance is $\approx 0.75$ dB and the mean CRT distance is $\approx 0.2$ dB.

\begin{figure*}[t!]
    \centering
    \subfloat[][Selected vehicular trajectory.]{\includegraphics[width=.45\textwidth]{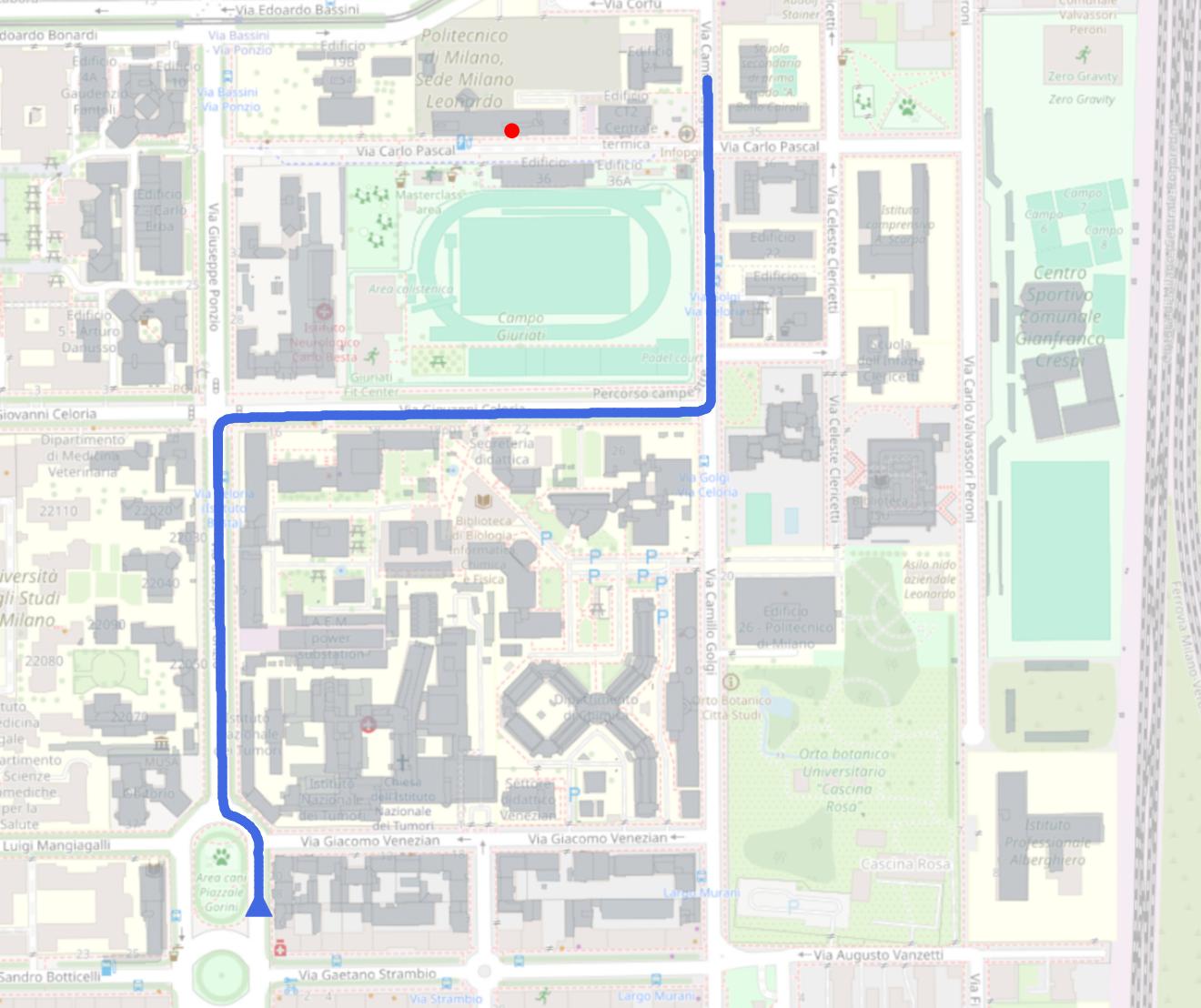}}
    \hspace{0.1cm}
    \subfloat[][Power distance (dB)]{\includegraphics[width=.515\textwidth]{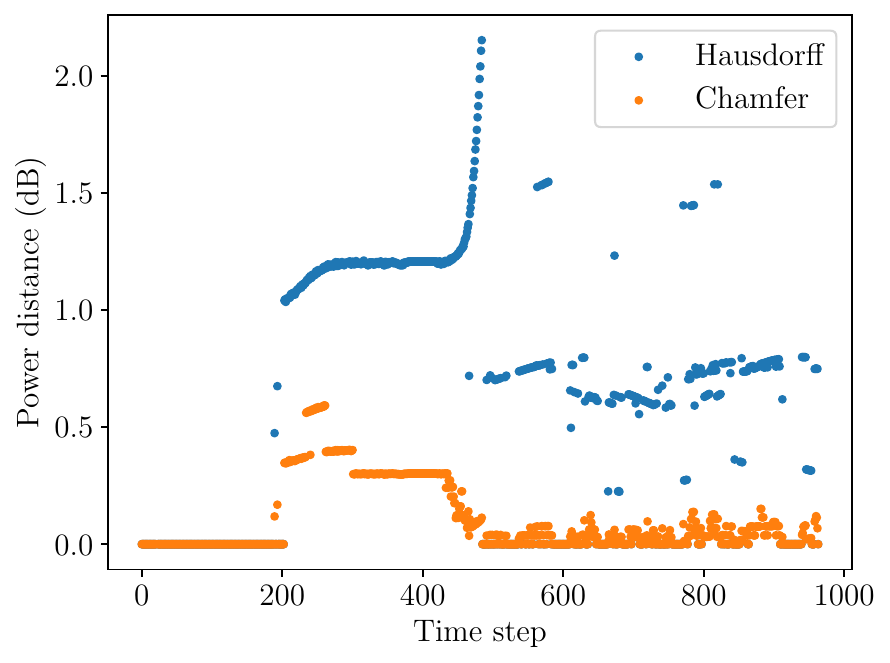}}\\
    \caption{Evaluated Hausdorff (HRT) and Chamfer (CRT) distances over power for a vehicular trajectory on the scenario enriched with building windows segmentation. In (a), the red point indicates the position of the BS, the vehicular trajectory is represented in blue, and the triangle on one of the extremes of the trajectory indicates its start.}
    \label{fig:windows_veh_results}
\end{figure*}

\textbf{Vehicular ray tracing simulations}: As for the parked vehicles scenario, we focus the vehicular analysis on a single trajectory representative of the observed distance patterns in the environment. We provide in Fig. \ref{fig:windows_veh_results} the results collected on the comparison performed on the selected vehicular trajectory. As for grid-based simulations, we focus on the power HRT and CRT distances, as the angular and temporal differences are minor owing to the considered modifications. We observed a maximum HRT power distance of $\approx 2.15$ dB and an average HRT distance of $\approx 0.55$ dB, while the maximum CRT distance is $\approx 0.59$ dB and the mean CRT distance is $\approx 0.11$ dB. For both HRT and CRT on rays' power, the distance presents patterns similar to the ones observed in the discussion in Section \ref{sec:results_parked}, mainly owing to the observation of the distance over Rx positions based on the sequential steps of a vehicular trajectory crossing the scenario. The observed pattern is consistent with the grid-based simulation in Fig. \ref{fig:windows_grid_results}. Indeed, the trajectory crosses areas corresponding to major differences between the compared scenarios, as highlighted by the patterns in both the HRT and CRT metrics.

For this comparative analysis, we examined a set of varied vehicular trajectories produced by the SUMO vehicular traffic simulator, observing patterns similar to the ones depicted in Fig. \ref{fig:windows_veh_results}. Given the density of the grid considered for grid-based simulations, the vehicular patterns were consistent with the distance values observed on grid. At the same time, the analysis over vehicular trajectories provides a dual representation that highlights the HRT and CRT distance variations on recurrent vehicular trajectories crossing the propagation environment.

\subsection{Discussion}

For both the considered scenarios, the proposed HRT and CRT metrics over power, temporal and angular ray features have highlighted the areas where corresponding ray tracing simulations differ, accounting for sets of paths with possibly different cardinality and providing distance measure based on the joint consideration of all the path features. The main differences were concentrated near the BS, with different patterns and different magnitudes depending on the environmental changes introduced in the scenario.

Both HRT and CRT distances were relevant for the comparison as they provided two different statistics in terms, respectively, of \textit{maximum} distance and of \textit{average} distance among the ray tracing features sets. As a result, HRT can be used to upper bound the differences among ray tracing simulations over different environmental conditions as computed over all the simulated rays, allowing to easily point out the emergence of at least one significantly different path configuration among a Tx and an Rx when the features of the environment are varied, as can be seen from Figs. \ref{fig:hausdorff_parked_grid_results} and \ref{fig:parked_veh_results}. On the contrary, CRT takes into account the properties of all the paths and results in increased robustness with respect to outliers. When the simulations differ only slightly, HRT and CRT can still qualitatively differ as HRT tended to \textit{saturate} to more uniform values for contiguous areas while CRT can still present more diversified spatial patterns across the scenario. This can be easily seen from Fig.~\ref{fig:windows_grid_results}, where the introduction of the windows on the buildings' facades slightly changes the power properties of the propagation rays while leaving the spatial and temporal features unchanged. A similar behavior can also be observed from Figs. \ref{fig:hausdorff_parked_grid_results} and \ref{fig:chamfer_parked_grid_results}, and from Fig. \ref{fig:parked_veh_results}, where the impact of parked vehicles modeling leads to higher HRT and CRT values while still showing relevant differences between the two metrics.

To examine the ray tracing dissimilarities across different levels of power for the simulated rays, a set of power thresholds may be considered to filter out less powerful rays. The HRT and CRT metrics can then be computed for each threshold to determine the radio propagation differences when different levels of importance are considered. As an alternative, the proposed metrics can be modified to weigh each ray contribution to the distance depending on its power. We notice that, although this path can be pursued to achieve a qualitative analysis, the use of a different weight for each path will produce the loss of interpretability in the temporal, power and angular features composing the proposed metric, as their measure units will be corrupted by the scaling process and by the nearest neighbor evaluation within the HRT and CRT metrics computation.

In this paper, the utility of the proposed metrics is focused on the analysis of the differences between the same Tx and Rx by varying only the environmental features and, therefore, the wireless propagation environment. We highlight that these metrics can be also useful to assess the spatial consistency of ray tracing simulations for neighboring sets of transmitters or receivers across a single scenario. In this case, the comparison can be performed between the ray tracing results associated to user equipment positioned nearby (e.g., within a predefined distance radius or divided into spatial clusters) and can provide meaningful insights on the channel's spatial variation.

Besides the provided analyses at 28 GHz, we performed analyses also at 7 GHz carrier frequency obtaining analogous results. While the simulations at 7 GHz showed higher coverage and a higher number of relevant paths per simulation---presenting differences that propagated further from the BS,---the observed patterns were similar to the ones proposed in this work, assessing the consistency of the metrics also for simulations carried out at lower bands.

\section{Conclusion}\label{sec:conclusion}

In this paper, we introduced the Hausdorff ray tracing (HRT) and chamfer ray tracing (CRT) metrics to compare the temporal, angular and power differences between two ray tracing simulations carried out on 3D scenarios featured by environmental changes. We considered a high-fidelity digital twin model of an area of Milan, Italy and we enriched it with two different types of environmental changes to produce two comparison scenarios over which to test the proposed metrics. We extended the base reference scenario by (i) including the presence of parked vehicles meshes in usual positions derived from satellite imagery of the area, and (ii) we segmented buildings facade faces in the scenario 3D meshes to separate the windows mesh components from the rest of the building. We performed grid-based and vehicular ray tracing simulations at 28 GHz carrier frequency on the base scenario and on the enriched environments by integrating the NVIDIA Sionna RT ray tracing simulator with the SUMO vehicular traffic simulator. Finally, we applied the proposed metrics to compare the base scenario with the ones containing the introduced environmental changes. Both the HRT and CRT metrics highlighted the areas of the scenarios where the simulated radio propagation features differ owing to the introduced environmental modifications. Moreover, the ray tracing simulation carried out over sets of Rx's located according to the vehicular simulations highlighted the patterns arising over HRT and CRT along realistic vehicular trajectories.

\section*{Acknowledgment}
The research has been carried out in the framework of the Joint Lab between Huawei and Politecnico di Milano. The windows segmentation work on the Digital Twin of the considered area of Milan, Italy has been performed by Zhonghui Liao as part of his M.Sc. thesis. The authors acknowledge the support of the Laboratorio di Simulazione Urbana ``Fausto Curti" of the Department of Architecture and Urban Studies of Politecnico di Milano in the provision of the highly detailed urban digital twin maps of an area of Milan, Italy used in this work.

\bibliographystyle{elsarticle-num}
\bibliography{Bibliography}

\end{document}